\documentclass{elsart}

\usepackage{psfig}
\usepackage{astron}

\newcommand{\be}{\begin{equation}}

\newcommand{\ee}{\end{equation}}

\newcommand{\Msun}{M_{\odot}}

\newcommand{\tamanio}{19.5cm}
\newcommand{\ptamanio}{12cm}
\newcommand{\pptamanio}{7.6cm}

\begin{document}

\begin{frontmatter}

\title{ Hydrodynamical simulations of galaxy properties: 
 Environmental effects}

\author[Madrid]{D. Elizondo\thanksref{DGICyT}}
\author[Madrid]{ G. Yepes\thanksref{DGICyT}}
\author[Alemania]{ R. Kates\thanksref{DGF}}
\author[EEUU]{ A. Klypin}
\address[Madrid]{Grupo de Astrof\'{\i}sica, 
Departamento de F\'{\i}sica Te\'orica C-XI, Universidad
Aut\'onoma de Madrid, Cantoblanco 28049, Madrid, Spain}
\address[Alemania]{ Astrophysikalisches Institut Potsdam, Potsdam, Germany}
\address[EEUU]{ Department of Astronomy, New Mexico State University,
Las Cruces, NM 88001, USA }
\thanks[DGICyT]{Work partially supported by DGICyT (Spain) under
project number PB93-0252} 
\thanks[DGF]{Work partially 
supported by a DFG fellowship (Germany)}

\begin{abstract} 

Using $N$-body+ hydro simulations we study relations between the local
environments of galaxies on $\approx 0.5 {\rm Mpc}$ scale and properties of
the luminous components of  galaxies. Our numerical simulations include
effects of star formation and supernova feedback in different
cosmological scenarios: the standard COBE-normalized Cold Dark Matter
model (CDM), its variant, the Broken Scale Invariance model (BSI), and 
a model with cosmological constant ($\Lambda$CDM).  The present time
corresponds to quite different stages of clustering in these three
models, and the range of environments reflects these differences.  In
this paper, we concentrate on the effects of environment on colors and
morphologies of galaxies, on the star formation rate and on the
relation between the total luminosity of a galaxy and its circular
velocity.  We demonstrate a statistically significant theoretical
relationship between morphology and environment. In particular, there
is a strong tendency for high-mass galaxies and for elliptical galaxies
to form in denser environments, in agreement with observations. We find
that in models with denser environments (CDM scenario) $\sim 13$ \% of
the galactic halos can be identified as field ellipticals, according to
their colors.  In simulations with less clustering
(BSI and $\Lambda$CDM), the fraction  of ellipticals is
considerably lower ($\sim 2-3 $\%). The strong sensitivity of
morphological type to environment is rather remarkable because
our results are  applicable to ``field''  galaxies and small groups.
Because of small  box size (5 Mpc)  we did not have large groups or
clusters in our simulations.

If  all galaxies  in our  simulations are included, we find a statistically 
significant dependence of the galaxy luminosity -  circular velocity 
relation on dark matter overdensity within spheres of radius $0.5$Mpc, 
for the CDM simulations. 
But if we remove ``elliptical'' galaxies from our analysis to 
mimic the Tully-Fisher relation for spirals, then no dependence
is found in any model.

\end{abstract}



\end{frontmatter}

\section{Introduction}

Most of our knowledge of the properties of 
galaxies derives from the light emitted by them, which in turn reflects 
the information contained in the baryonic component.  
In order to explain the origin of these properties, it is
necessary to model  the dynamics of the baryonic gas coupled 
gravitationally to the dark matter component.  
Previously, galaxy properties were studied using N-body simulations for
the dark matter, and assuming different hypothesis to ``extract''
information for the baryonic component, e.g. linear biasing
\cite{kaiser} or constant M/L.  Improvements in
computational  techniques and resources have made it possible to  include  gas
dynamics in N-body simulations \cite{evrard,cencen,kagun}.

In addition to purely gravitational and hydrodynamical effects, it is
essential to include complex local processes that affect the energy
budget of the gas, such as radiative cooling or photoionization heating
by an UV background.  It is also important to consider effects of
stellar formation and evolution and feedback processes produced by the
explosions of supernovae.  Self consistent modeling of all these
physical processes in cosmological simulations is crucial for attaining
insight into the generation and evolution of galaxies within a given
cosmological scenario.\footnote{See e.g. Yepes \cite{casa} for a recent
review of the inclusion of baryonic physics in cosmological simulations}.
However, numerical simulations invariably include various 
approximations and simplifying assumptions regarding the relevant 
physical processes; these assumptions 
need to be verified. In recent years there has been a
tremendous accumulation of observational data shedding light on the
underlying properties of galaxies. The observations also allow us
to constrain the assumptions in numerical modeling of the 
complex baryonic processes involved in galaxy formation and evolution.

As an alternative to the use of direct numerical simulations of the 
most important relevant physical phenomena, much of the effort
toward modeling relationships between environment and observational
properties of galaxies up to now has been focussed on "semi-analytical"
approaches to galaxy formation and evolution
\cite{whfr,kauff,lccl,semianallll,heyl,baugh}.  
However, although quite a bit has been learned from semi-analytical
models, there are certain processes that they are simply not 
designed to incorporate. In particular, the ability to take nonlinear
hydrodynamical processes and their interactions with other physical processes
properly into account is limited. Hydrodynamical effects could in principle 
be implemented in a semi-analytical approach, but this would require some
kind of prior knowledge to identify the most important consequences
and insert their effects correctly by hand. 
It is not clear whether the effort required
to do so is less than that required to simulate the whole problem
including hydrodynamics self-consistently right from the start.

In this article, we study environmental effects on observable 
galaxy properties such as colors, morphology, star
formation rates, and magnitude-circular velocity relations by means of
numerical simulations. 
For these simulations, we have adapted and applied the N-body (PM) + 
Eulerian hydro code ({\em Piecewise Parabolic Method\/}
 \cite{colella}) reported in Yepes, et al. \cite{gus}, (YK$^3$). 
As described below, this code is designed to model those
processes generally thought to be most relevant 
at the scales we are considering. 
The universe is modeled as a 4-component medium
consisting of dark matter, stars, ``hot" or ambient gas, and cold
clouds. Star formation depends on the local density of cold gas
clouds and is regulated by the supernova feedback loop. In particular, 
star formation
is modeled by converting cold gas clouds at a certain rate into
discretized particles, each of which may be idealized as representing a
``starburst."  Stellar population synthesis models \cite{bruzual}
 are then used to derive the luminosities and colors of the
resulting galaxies. 

To add more breadth to our analysis 
of environmental effects, it is useful
to consider the evolution of this 4-component medium 
in the context of several rather different 
cosmological scenarios: 
\begin{enumerate}
\item  Standard unbiased Cold Dark Matter model (CDM).  
\item  Model with a cosmological constant ($\Lambda$CDM). 
\item  Broken Scale Invariance (BSI) \cite{bsi95}. 
\end{enumerate}

These distinct scenarios allow us to compare how different
growth rates in cosmological models (e.g., CDM vis. $\Lambda$CDM) and
different epochs of structure formation (CDM vis. BSI) affect the final
properties of galaxies. 
However, since it is difficult to control for 
confounding factors such as the influence of the box size, this paper is not 
intended to distinguish a preferred cosmological model. 

The paper is organized as follows. Section   
\ref{sec:simu}
briefly reviews the hydrodynamical model used,
describes the halo identification algorithm
and the technique for assigning luminosities to these halos
and discusses the characteristics of the numerical experiments.
Section \ref{sec:dis} 
describes the general properties of  
the distribution of the four components of 
matter in the simulations. Section  
\ref{sec:env} summarizes the quantitative analysis of
the effects of environment on galaxy properties.  The conclusions
are given in Section \ref{sec:conclu}.

\section{Simulations}
\label{sec:simu}

\subsection{Description of the hydrodynamical
model} \label{sec:hydromodel}

The code which we have applied to this problem has been designed and tested
(YK$^3$) to provide a 
phenomenological  description 
of the interaction between the gas and the stellar
component, simulating the physical processes  
most relevant for galaxy formation:

(i) 3D hydrodynamics (including a shock capturing technique),
(ii) the dynamics of the interstellar medium, modeled as a two-phase 
medium
(ii) radiative and Compton cooling of the gas,
(iii) star formation and heating of the gas by supernovae explosions (''supernova 
feedback'').

Here, we summarize the equations governing the cooling
of the gas,  star formation, and the effects of supernovae in a
slightly simplified form.  The main characteristics of the
code were discussed in detail in YK$^3$.

The code models two mechanisms for transferring gas from the hot
component with density $\rho_h$ to the cold component with density
$\rho_c$: 
The latter (``cold clouds'') are intended
to represent structures where star formation occurs, 
such as Giant Molecular Clouds in
our galaxy.
If the gas temperature in a cell $T_h$ drops below $T_{lim}=2 \times
10^4$ K and the density of the gas exceeds a threshold, all the hot gas in
the cell is transferred to the cold component, thus enabling star
formation. The condition on the gas density is 
expressed in the form $\rho_{\rm gas}>
{\mathcal D} \; \Omega_B\rho_{cr}$, where $\Omega_B\rho_{cr}$ is the
mean density of the baryons in the Universe.  
The adjustable parameter ${\mathcal
D}$ is introduced \cite{muka} in order to take into account heating
of the gas by an ionizing ultraviolet background due to quasars and
active galactic nuclei \cite{giroux,petit,mu}:  at lower densities
the gas is significantly heated by the UV flux and cannot form
stars. Here, we took ${\mathcal
D}=100$. This mechanism also prevents star formation outside of 
galaxies.

Another mechanism for formation of cold gas clouds is by means of the
thermal instability \cite{nube2}.
The temperature range for production of cold clouds via thermal  
instability is taken to be $T_h < T_{inst}$, where in this paper
$T_{inst} \sim 5 \times 10^5$ K.  To compute the rate of mass increase
in cold clouds, we suppose that all of the energy emitted by the hot
gas is really lost by the gas which goes from hot to cold. If we
suppose in addition that the gas is ideal and cannot cool below
$T_{lim}$, then we obtain the expression

\be
\left( \frac{d \rho_{h}}{dt} \right) = - \left( \frac{d
\rho_{c}}{dt} \right) =  - C \frac{\Lambda_r (\rho_{h},
T_{h})}{\gamma \epsilon_{h} - \epsilon_{c}}, 
\ee
where $\gamma$ is the ration of specific heats 
of an ideal gas, $\epsilon$ is the thermal energy per unit
mass of gas, $\Lambda_r$ is the cooling rate due to radiation, 
and the factor $C=1-10$ is introduced to
characterize the uncertain 
effects of unresolved substructure. Nevertheless, at it was shown in
$YK^3$, the results are not sensitive to  the value of the $C$ parameter. 

Because of the frequent exchange of mass between the hot and cold gas
phases, it is reasonable to consider the hot gas and cold clouds as 
{\it one}
fluid with rather complicated chemical reactions going on within it.
The two components are assumed to be in local pressure equilibrium
so that 
$P_{\rm gas}=P_{c}=P_{h} \equiv (\gamma -1) u_{h}$, $u_{h} 
 = \rho_{h} \epsilon_{h}$, where $u$ is the internal energy
 per unit volume and  $P_{\rm gas}$ is the internal gas pressure.

Star formation is assumed to occur only in cold clouds.  
Stars with masses exceeding  
$(10-20) M_{\odot}$  undergo supernova explosion
within a single simulation timestep.  If we suppose that
the stars that explode on this short timescale
constitute a fraction $\beta$ of the total, the rate
of formation of long-lived stars becomes 
\be
\frac{d \rho_*}{dt} = \frac{(1- \beta) \rho_{c}}{t_*},
\ee 
where $t_*=10^8$yr is the time scale for star formation.
The value of $\beta$ is sensitive to the 
shape of the initial mas function (IMF), especially the lower
limit.  In our simulations we have taken $\beta=0.12$ 
corresponding to the Salpeter IMF. In view of the
 uncertainties in supernova energy, the energy input
 due to lower-mass stars, $(7-10) M_{\odot}$, is not included here.
The evaporation of cold clouds 
in the interstellar medium due to supernova explosions 
is incorporated into the code by supposing that the total mass of
cold gas transferred to the hot gas component is not just
the mass in the star itself, but a factor of $A$ times
the supernova mass. 
Each $1\Msun$ of supernovae dumps
$4.5\times 10^{49}$ ergs of heat into the interstellar medium  and
evaporates a mass $A\cdot\Msun$ of cold gas. This ``supernova
feedback parameter'' $A$
could well depend on local gas properties such as density and chemical
composition. 
Previous results ($YK^3$)  point  to
a large value of $A$. An especially sensitive test is provided by the
Tully-Fisher relation \cite{nosotros}.  Here, we will assume 
 $A$ to be constant and large ($A=200$)
resulting in low efficiency of
converting cold gas into stars. The  mass transfer between different
components, due to evaporation,  is defined by:
\be
\left( \frac{d \rho_{h}}{dt} \right)_{evap} = - \left( \frac{d
\rho_{c}}{dt} \right)_{evap} = \frac{A \beta \rho_{c}}{t_*}
\ee
Another important effects of supernovae explosions is the injection of
metals to the  surrounding gas,  which modifies  its  cooling
properties. In order to incorporate the effects of metallicity in our
code, we assume solar abundance in regions where previous star
formation has occurred. Otherwise,  the region is not considered to be
enriched with metals, and cooling rates for a gas of primordial
composition are assumed. 

Because supernova feedback regulates star formation efficiency in halos, the
value of $A$ would be expected to play an important role in the observational 
consequences of our results.  Hence, comparison with observations
should provide useful information on supernova feedback.  
For this  reason, several of the
simulations reported in Table (\ref{tab:para}) have been rerun with
different values of $A$, in order to study the effects of feedbacks on
the observational properties of galactic halos. 
 In previous work, we have paid particular
attention to the  effects of feedback on  
the characteristics of the magnitude-circular velocity (Tully-Fisher)
 relation and on the  faint end of the luminosity functions in different
 photometric bands \cite{yopots}.  We have seen that supernova
 feedback is responsible mainly for the determination of the slope of
 the Tully-Fisher: low $A$ values (large reheating) makes the low
 circular velocity  halos  fainter while hardly affecting the
 halos with large circular velocity. The scatter of the relation is
 reduced when low $A$ values are assumed, as compared with large $A$
 values. In the case of no feedback ($A=0$), we get slopes  and scatter 
 that are inconsistent with the observational Tully-Fisher. Also, when
 no feedback is assumed, we get a faint-end luminosity function which
 is too steep (too many faint galaxies)
 as compared with the most recent estimates for the $B$
 and $K$ bands. More detailed information on this analysis can be
 found elsewhere \cite{nosotros}. Here, we will focus our study on 
 a possible dependence  of the Tully-Fisher relation  
(see {\S} \ref{sec:tf}) on environment.

\subsection{Galaxy finding algorithm}

Galaxy catalogs at selected
redshifts were constructed from the simulation data as described in
YK$^3$. Briefly, local maxima of the dark matter density field 
are first identified on the grid.  
We next compute the mass inside spheres of 2-cell radius ( 78 kpc) 
centered at the center of mass of the local particle distribution.  If
the mass inside such a sphere exceeds the mass within a sphere of
overdensity 200, we mark the 2-cell sphere as a dm halo. 
Otherwise, we repeat the process  
for a sphere of 1-cell radius ( 39 kpc). 
If the test succeeds, the matter within the sphere is assigned to the halo. If the 
test again fails, the local maximum is not considered to be a halo. Some dm 
halos have zero luminosity.  
In what follows, we use the term ``galaxy" to refer to a halo with
nonzero luminosity.

\subsection{Assignment of magnitudes and luminosities}

In order to assign luminosities and colors to the halos in which stars have formed,
we need to know the spectral energy distribution (SED)
$S(\lambda,t)$ for each
halo as a function of time.  This is possible since for each galaxy we have a
list of starbursts and, associated with each, its mass and time of formation.
These two parameters are sufficient for calculating the SED assuming an
appropriate model for stellar population synthesis.  
Among the numerous stellar
population synthesis models that have been proposed (see YK$^3$ for references
and discussion)
we have chosen the model of Bruzual \& Charlot (1993), which describes the time
evolution (between 0 and 20 Gyr) of SED's for a burst of star formation under
conditions of solar metallicity and a Salpeter initial mass function,
 in accordance with the assumptions of the YK$^3$ code. 

Using this stellar population synthesis
model, we compute the 
SED of each galaxy according to 

\be
\label{SED2}
S(\lambda,t)= \sum _{\tau_i} \Phi(\tau_i) {\mathcal F}(\lambda,t-\tau_i),
\ee
where  $\Phi(\tau_i)$ is the mass of stars in the halo  produced at timestep
 $\tau_i$ ,  and
${\mathcal F}(\lambda, t)$ is the SED due to a starburst of 1 $\Msun$
after an evolution time $t$.

To obtain the
absolute luminosity $L_f(t)$ in any given band, 
we convolve $S(\lambda,t)$ with the appropriate 
filter response function
$R_f(\lambda)$. Combining the $L_f(t)$, we then obtain the evolution of
the color index of the 
numerical 
galaxy.  (Of course, the observed color
of a real galaxy would be influenced
by additional factors such as the interaction of starlight
with the surrounding plasma).  We have calculated the luminosities and colors
of the galaxies in the various filters comprising the Johnson UBVRIK system.

\subsection{Parameters of the  simulations} 
\label{subsec:simupar}
Regarding our simulations as numerical experiments, our 
goal was to obtain a sufficiently large 
sample of ``numerical galaxies'' to permit reliable, i.e., statistically
significant comparisons with observational quantities.  To this end, a
set of 11 simulations were performed for each of the CDM, $\Lambda$CDM
($\Omega_\Lambda = 0.65$), and BSI models.  COBE normalization was
taken and baryon fractions were compatible with nucleosynthesis
constraints \cite{smith} ($\Omega_B=0.051$ for BSI and CDM,
$\Omega_B=0.026$ for $\Lambda$CDM ).

Due to limitations on computational resources affecting both the number
of particles and the number of cells, the selection of a simulation
volume requires a compromise among various considerations. On the one
hand, we need to resolve scales on the order of the size of a galaxy,
implying high resolution, but on the other hand the larger the volume,
the more the simulations will be representative of the range of
environments possible in the cosmological scenarios.  Based on   
general considerations and on our experience with the performance of
the code, we have taken a size of 5~Mpc but with different Hubble
constants given by $h=0.7$ for the $\Lambda$CDM model and $h=0.5$ for
the CDM and BSI simulations.
In what follows, all lengths and masses are
expressed relative to the corresponding Hubble constant of particular
model (i.e., the units are scaled to ``real'' megaparsecs, and there are 
no $h$ terms). 

Due to the small box size of our simulations, needed to properly
resolve galaxies, most of the large-scale power of the density
fluctuations is absent. In order to characterize the effects of the missing 
longer-wave fluctuations on the structures that are formed inside our
boxes, we compute the linear theory mass variance,
\be
\sigma^2_L(R)= \int_{2\pi/L }^{kmax} e^{-k^2R^2}P(k) k^2 dk
\ee
with a Gaussian filter of $R=0.5$ Mpc for boxes of  $L=5$ Mpc 
and for boxes of 
$L=10$ Mpc. The resolution is the same in both cases (same $kmax$).
Here, we use a CDM power
spectrum (Fig \ref{fig:esp2}), but the results will be similar for other
models because the spectra are basically parallel in the range of
wavenumbers we are interested in. The ratio of mass
variances is $\sigma_{10}(R=0.5)/\sigma_{5}(R=0.5) = 1.12$. This linear estimate
suggests that on scales of 0.5 Mpc, 
which is the one we will use to study the effects of environment, the 
effects of the missing power will not be very important ($\sim 12$\%). 

 The
simulations reported were performed in two SGI Power Challenge 
supercomputers at the  CEPBA (Centro Europeo de Parelismo de Barcelona). 
The main characteristics of the simulations are given  in Table 
(\ref{tab:para}).

\begin{table}
\caption{Parameters of the simulations.\label{tab:para}}

\begin{tabular}{lcccc}         
  \hline  \hline Parameters & CDM & $\Lambda$CDM &$\Lambda$CDM& BSI \\ 
  \hline 
 $\Omega_{\Lambda}$ & 0 & 0.65& 0.65 & 0 \\
 $\Omega_{B}$  & 0.051 & 0.026& 0.026 & 0.051 \\ 
$\Omega_{dm}$ & 0.949 & 0.324& 0.324 & 0.949 \\
 $h$                      & 0.5 & 0.7 & 0.7 & 0.5 \\
  Simulation Box at $z=0$, ($h^{-1}$Mpc) &2.5 & 3.5& 3.5 & 2.5 \\
 Cell size at $z=0$, ($h^{-1}$kpc) & 19 & 27& 13.7 & 19 \\
  Number of cells & $128^3$ &  $128^3$ &$256^3$ & $128^3$ \\ 
  Number of realizations & 11 & 11 & 1& 11 \\ 
 Mass Resolution for dark matter ($h^{-1}M_{\odot}$) & 
           $2\times 10^6$ & $1.3\times 10^6$ &$1.6\times 10^5$& $2\times 10^6$\\
 Total  Number of galaxies & 447 & 442 & 50 & 649 \\
 Total number of bright galaxies
  ($M_B \leq -16$) & 73 & 91 & 8 & 144 \\ \hline 
\end{tabular}
\end{table}

The 11 simulations carried out for each model correspond to a total
simulated comoving volume of 1375 Mpc$^3$.  The total number of {\em
numerical} galaxies generated is of the order of $\sim 450$ for each
model. This quantity of data permits construction of a reasonably large
sample (data base) suitable for carrying out statistical analyses
sensitive to the effects we would like to study and comparing with
the appropriate observations. 

\subsection{Effects of resolution} 
\label{subsec:resolution}

\begin{figure}
    \centerline{\psfig{file=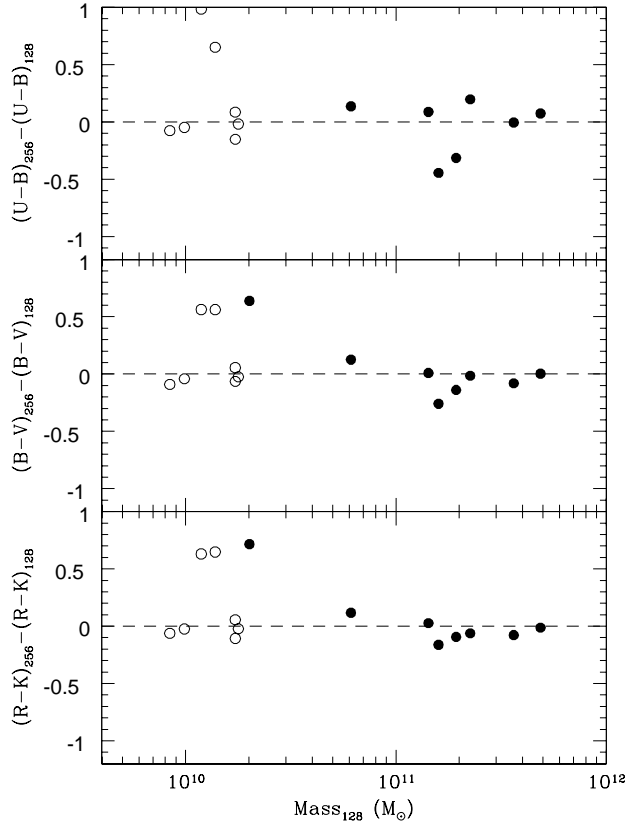,height=\ptamanio}}
    \caption{Variation of   the  color indices  ($U-B$, ~$B-V$ and
      $R-K$) of the galaxies  found in one  $\Lambda$CDM simulation run with 
   two
      different resolutions: $128^3$ and $256^3$, as a function of the
      halo mass. Filled circles represent galaxies brighter than 
      $M_B \leq -16$. Open circles represent faint galaxies ($M_B > -16$).
}
    \label{fig:resol}
\end{figure}

In any cosmological hydrodynamical simulation, limitations on resolution
give rise to a smearing of
internal structures, which constitutes a potential source of error. 
The simulations reported here are no exception, 
because with a spatial resolution of 10~kpc to 40~kpc, we clearly do not
resolve the internal structure of a galaxy. Limits on spatial resolution
also degrade the time resolution of star formation, which is also limited for other 
reasons. Time 
resolution, though not often discussed, is of comparable importance to 
spatial resolution, because moderate but recent bursts of star formation that
are not properly resolved (e.g., attributed to earlier times)
can have a large effect on colors (see Fig. \ref{fig:resol}).

Having said all this, we note that for the most part global parameters
(such as the total luminosity or maximum of the rotational velocity)
turn out to be rather insensitive to resolution. 
This ``robustness'' was confirmed by
comparing results of simulations made with the same initial conditions,
but run at different resolutions. Some tests concerning 
the effects of the resolution were 
presented in YK$3$ and in Elizondo {\em et al} (1998),
 but due to the importance of this
issue, we discuss some of them here: 

In order to test  the effects of spatial numerical resolution on the 
properties of the halos, we have run one simulation for
the $\Lambda$CDM model, with the same box size and parameters as for 
the 11 simulations of the $\Lambda$CDM model given in Table 1, 
but with $256^3$ particles
and cells (i.e. 19.5~kpc comoving and $2.3\times 10^5 M_\odot$
mass per particle). We  then reran this simulation  at  lower
resolution with $128^3$ particles and cells. The initial particle  
distribution was generated according to the Zeldovich 
approximation. The displacement field used for the  $128^3$ grid was
generated by aggregating the displacement field 
over the 8 nearest-neighbor cells with respect to the $256^3$ grid. 
The high-resolution simulation produces about $\sim 25$\% more galaxies,
but the excess comprises primarily faint galaxies. 
The individual massive halos remaining at the end of evolution in both
simulations could be identified one-to-one,
which made  possible a detailed and reliable comparison of the effects of
resolution in the final observational properties of the halo
distribution.



\begin{figure}
    \centerline{\psfig{file=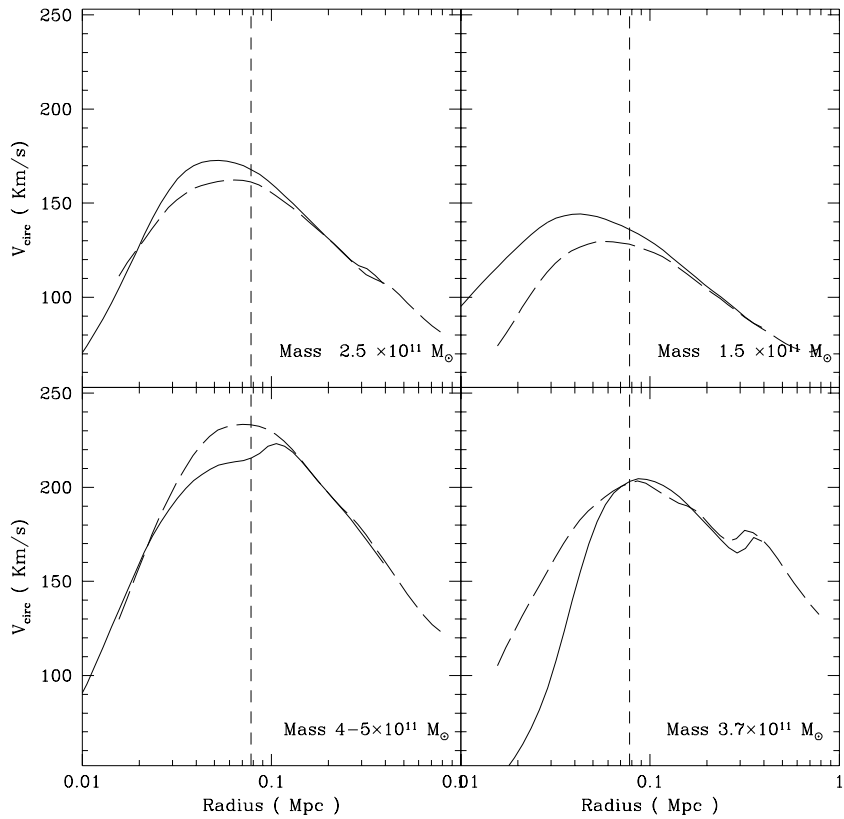,height=\ptamanio}}
    \caption{Effects of resolution on the circular velocity profile of
       the 4  most  massive  galaxies found in one  $\Lambda$CDM simulation run with 
   two different resolutions: $128^3$ and $256^3$ (column 3 of Table
   \protect\ref{tab:para}).  Solid line is the circular velocity for the
   high-resolution simulation.
   Dashed lines correspond to the halos found in the same
  simulation  but with lower resolution. The vertical dotted
 line represents the limiting radius for these galaxies  according to our
 galaxy finding algorithm.}
    \label{fig:testv}
\end{figure}

In Figure (\ref{fig:testv}) we plot the circular velocity profile
(($G M(r)/r)^{1/2}$) for the 4 biggest halos found in the $\Lambda$CDM
simulations mentioned above. As can be seen, estimates of the circular 
velocity at the limiting radius 
for these halos  differ by less than 10\%.

The ``robustness'' seen here can be understood if we consider
the evolution of the gas falling on a galaxy 
containing hot gas. For a large enough galaxy
(where the limit for ``enough'' does depend on resolution) the gas cannot be
pushed back too far even if the galaxy has a high star formation rate (SFR).
Because the gas density close to a galaxy is high, it appears that the
cooling time is always rather short (on cosmological scales). As the
result, the gas keeps falling back on to the galaxy, cools even faster, and
gets converted into stars. Because 
in our model the whole process of star formation 
from the gas (which determines the SFR and thus the luminosity) is 
regulated mostly by the rate of gas infall and by feedback (neither of which 
depend strongly on resolution), the luminosity and colors of the galaxy are
remarkably
unbiased with respect to resolution changes. 
However, there are clear limitations to this weak dependence
on resolution: Small
galaxies (in our case $M_{total}< 3\times 10^{10}\Msun$) are indeed sensitive
to the resolution, because in their case the depth of
the potential well is not accurately modeled.



In order to estimate the effects of numerical  
resolution on the morphological assignment 
  used in this paper (see section \ref{sec:mfenv}),
  we have computed the color indices
($U-B$,~$B-V$ and  $R-K$) of the halos for the $\Lambda$CDM  simulation  run
with $256^3$ particles and cells 
and those from the same halos found in the simulation rerun with 
lower resolution ( $128^3$ particles and cells).  
Fig. (\ref{fig:resol}) shows the variation of the color indices  as a
function of the mass of halos found in the lower-resolution simulation
 From this
figure we can conclude   that the color indices  of massive halos
(i.e. $M \geq  3\times 10^{10} M_{\odot}$) are not very sensitive to 
resolution. Some of them do have bluer colors (specially $U-B$)
 in the high-resolution simulation  
due to a recent burst of star formation in those halos which did not occur 
in the low-resolution simulation. For these halos,
the morphological assignment according to colors  would have been affected
by resolution. This is the sort of effect that would have been expected and
should be kept in mind when interpreting the results. 
Nevertheless, the main conclusions of our qualitative
study of the dependence of morphology on environment appear not to be 
seriously affected.

\section{General Properties of the Simulations}
\label{sec:dis}

We begin our description of the simulation results in our
CDM, $\Lambda$CDM and BSI scenarios with some qualitative aspects of the
simulated gas and dark
matter distributions. 


\begin{figure}
\centerline{ \psfig{file=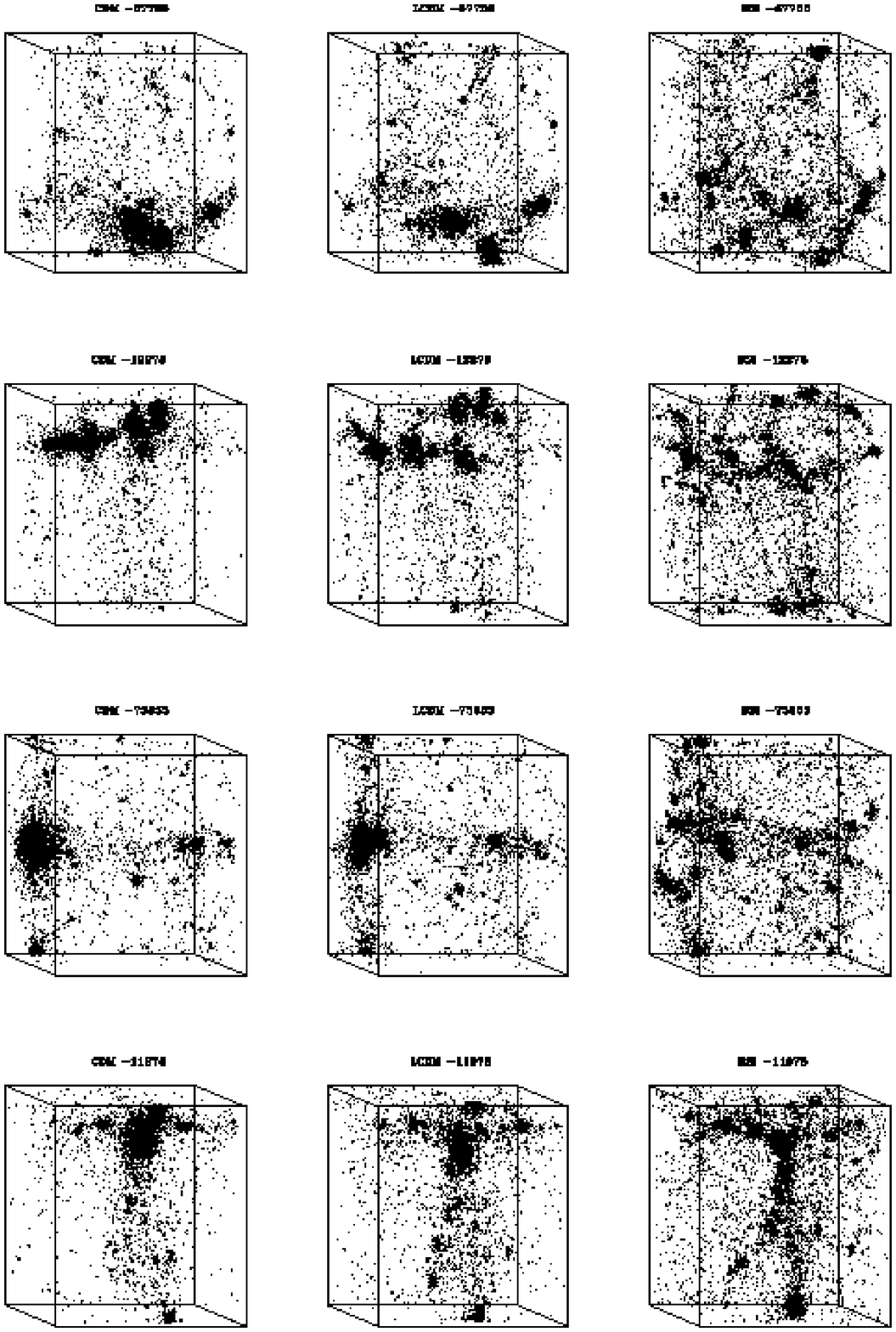,height=\tamanio}}
\caption{3-dimensional distribution of  10\%  of the dark matter
  particle at $z=0$ for 4 of the 11  realizations. The first column represents 
CDM realizations, the second $\Lambda$CDM (LCDM)
and the third is for the  BSI.  Each group of  realizations (CDM,
$\Lambda$CDM and BSI) was performed with the same random phases.
The random seed chosen  for the generation of the phases  is
indicated in each box. The box size is 5 Mpc} 
\label{fig:3ddark}
\end{figure}



\begin{figure}
\centerline{ \psfig{file=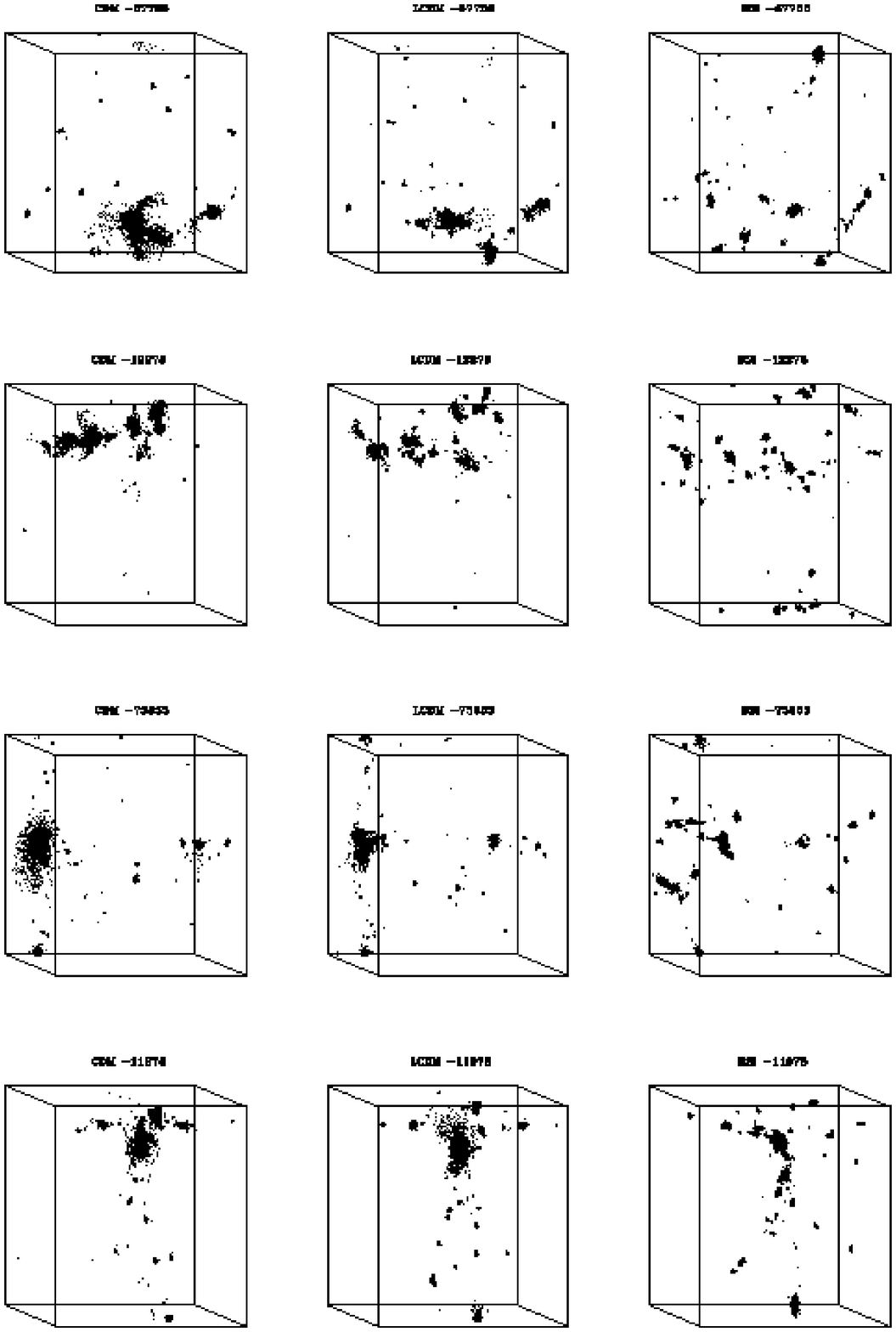,height=\tamanio}}
\caption{Same as  in Fig. \protect\ref{fig:3ddark} but for the star
particle distribution.} 
\label{fig:3dstars}
\end{figure}

\subsection{Dark matter distribution}
\label{sec:darkdistr}

\begin{figure}[th]
    \centerline{\psfig{file=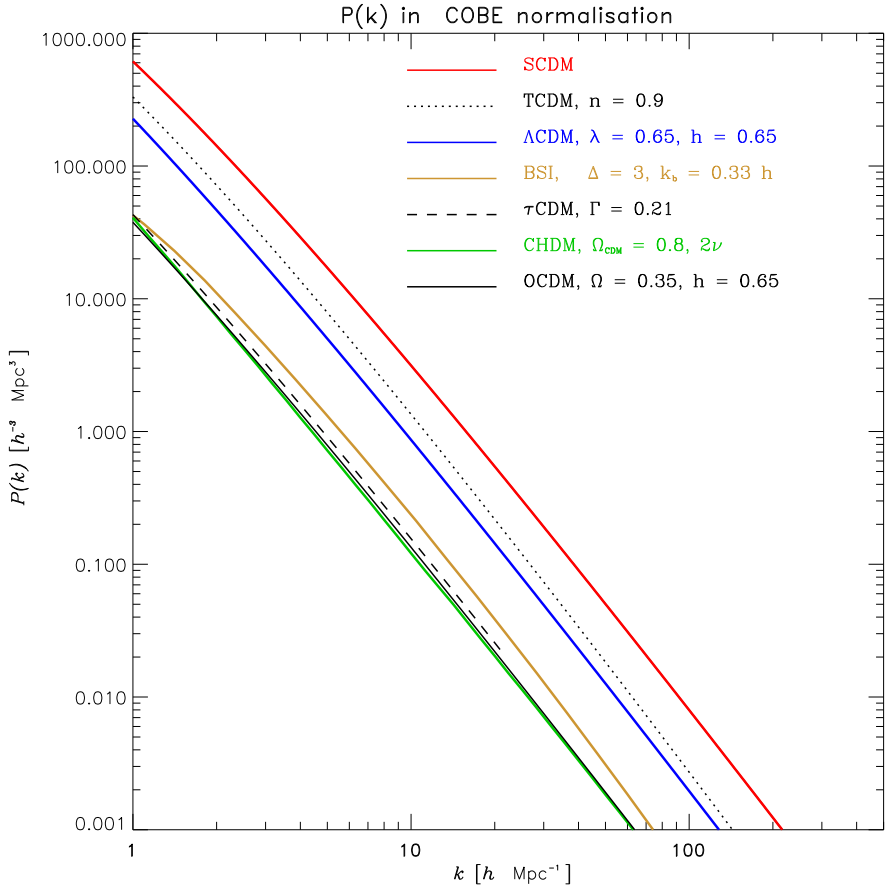,height=\ptamanio,angle=0}}
    \caption{Power spectrum for eight of the most {\em popular} 
cosmological scenarios, at the  scales relevant for our analysis. 
All of them are COBE normalized.} 
    \label{fig:esp2}
\end{figure}

Fig.  (\ref{fig:3ddark}) and (\ref{fig:3dstars}) show the distribution of dark
matter particles at redshift $z=0$ for each realization of the three scenarios
considered.  (Since only 10\% of the particles have been plotted, some of the
structures appear slightly washed out).  One sees that both CDM and
$\Lambda$CDM have given rise to numerous galaxies, with one predominant massive
($M \sim 10^{12} M_{\odot}$) galaxy generated by merger of different halos
during the course of its evolution.  In both cases, one observes a significant
number of less massive halos distributed in a transitory (but still present)
filament.  In many simulations, the most massive galaxy is accompanied by
neighboring halos with a mass of the order of $M \sim 5 \times 10^{10}
M_{\odot}$, which are destined to be absorbed by the main galaxy due to the
intense gravitational attraction.  A general trend in our results is that
the massive halos in CDM are the product of a larger number of collapses than the
''corresponding'' structures in 
in $\Lambda$CDM. This difference is especially evident in the four CDM realizations
 shown in Fig. \ref{fig:3ddark}) and (\ref{fig:3dstars}),
 in which the massive galaxy is just merging with a
neighbor or has just devoured one, whereas in the corresponding $\Lambda$CDM
realizations the galaxies are still approaching.  In the same manner,
clustering is more pronounced in CDM than in the other scenarios.  
This result might at
first seem puzzling in $\Lambda$CDM, since all other things being equal, the
time available for evolution tends to be longer in models with a cosmological
constant.  However, our choice of parameters implies a similar evolutionary
time for all the models under consideration, and hence the degree of clustering
depends mainly on the initial power spectrum.  As seen in Fig.
(\ref{fig:esp2}), the CDM spectrum has a larger amplitude at the scales
considered than $\Lambda$CDM; this difference is sufficient to explain the
disparity between the two models.


In the BSI models, the differences are even greater.   The distribution of
matter appears much more diffuse and the filaments more evident 
(see simulation 11978). The number of halos is much greater, but they 
are less strongly clustered than in the other two models (see 12278). 
The larger halos generated in 
BSI generally have lower masses than the corresponding ones in CDM and
$\Lambda$CDM. It is evident that fewer mergers of halos have occurred and that 
the halos are distributed more evenly within the simulation volume.
This property may also be attributed to the  lower amplitude of the fluctuation
spectrum.  Nonetheless, there is a systematic relation in the evolution 
in CDM and BSI, since the spectra are
essentially parallel in the dynamical range studied here.  Hence, aside from 
an overall delay in the dynamical evolution of
BSI models relative to the corresponding CDM realization, there is a great
similarity in those dynamical properties and structures of the two
models which depend 
mainly on the dark matter distribution (clustering, matter
distribution, masses, etc.).   
Nonetheless, as will be seen in what follows, there are no
correspondingly simple 
relations characterizing those properties which depend 
mainly on the baryonic component.

The above discussion concerning the  similarity of BSI and CDM power spectra 
hardly applies to the $\Lambda$CDM model.
The observable properties, especially with respect to the 
baryonic component in $\Lambda$CDM and CDM, are even more 
distinct due to the differences in the quantity of dark and baryonic 
matter, in expansion rates, and in the Hubble parameter.

\subsection{Characteristic evolutionary stages of
dark and stellar matter in the scenarios}



\begin{figure}
\centerline{ \psfig{file=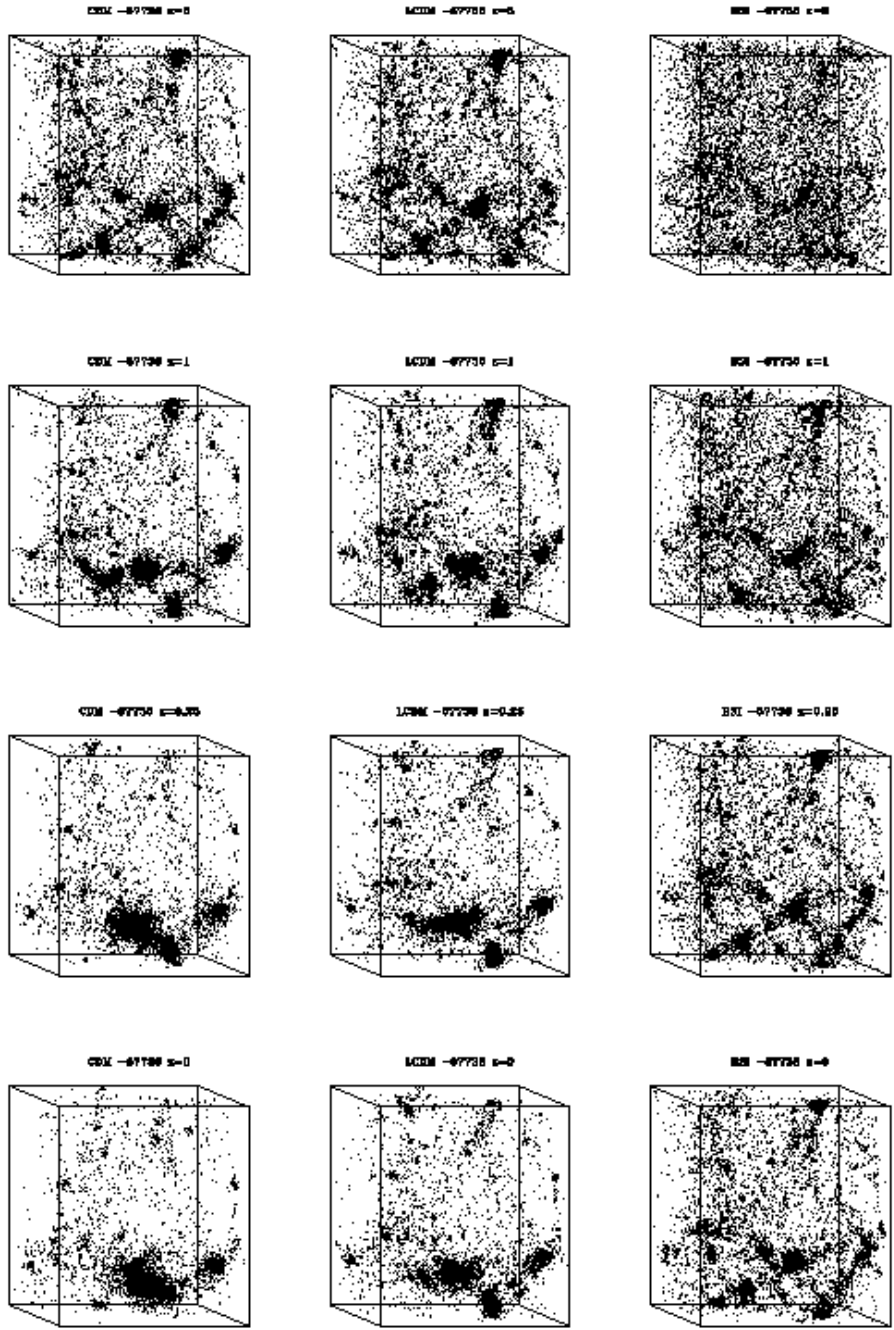,height=\tamanio}}
\caption{Redshift evolution,  from  $z=3$ to $z=0$, 
 of the dark matter particle distribution
for the realization 67736 in   the three cosmological scenarios. 10\%
of dark particles are shown.} 
\label{fig:3devoldm}
\end{figure}

Numerical simulations offer a useful opportunity for studying how the distinct
properties of halos evolve.  Some qualitative insight into the evolution of the
particle distribution is provided by  Fig   (\ref{fig:3devoldm}).
 The dark matter evolution from $z=3$ to $z=0$ is illustrated for the
realization 67736 of all three scenarios (to be studied in more detail below).
In these figures, one sees that CDM and $\Lambda$CDM follow a similar evolution
up to about $z \sim 1$.  Henceforth, a greater rate of collapse is observed in
CDM. 
\begin{figure}
\centerline{ \psfig{file=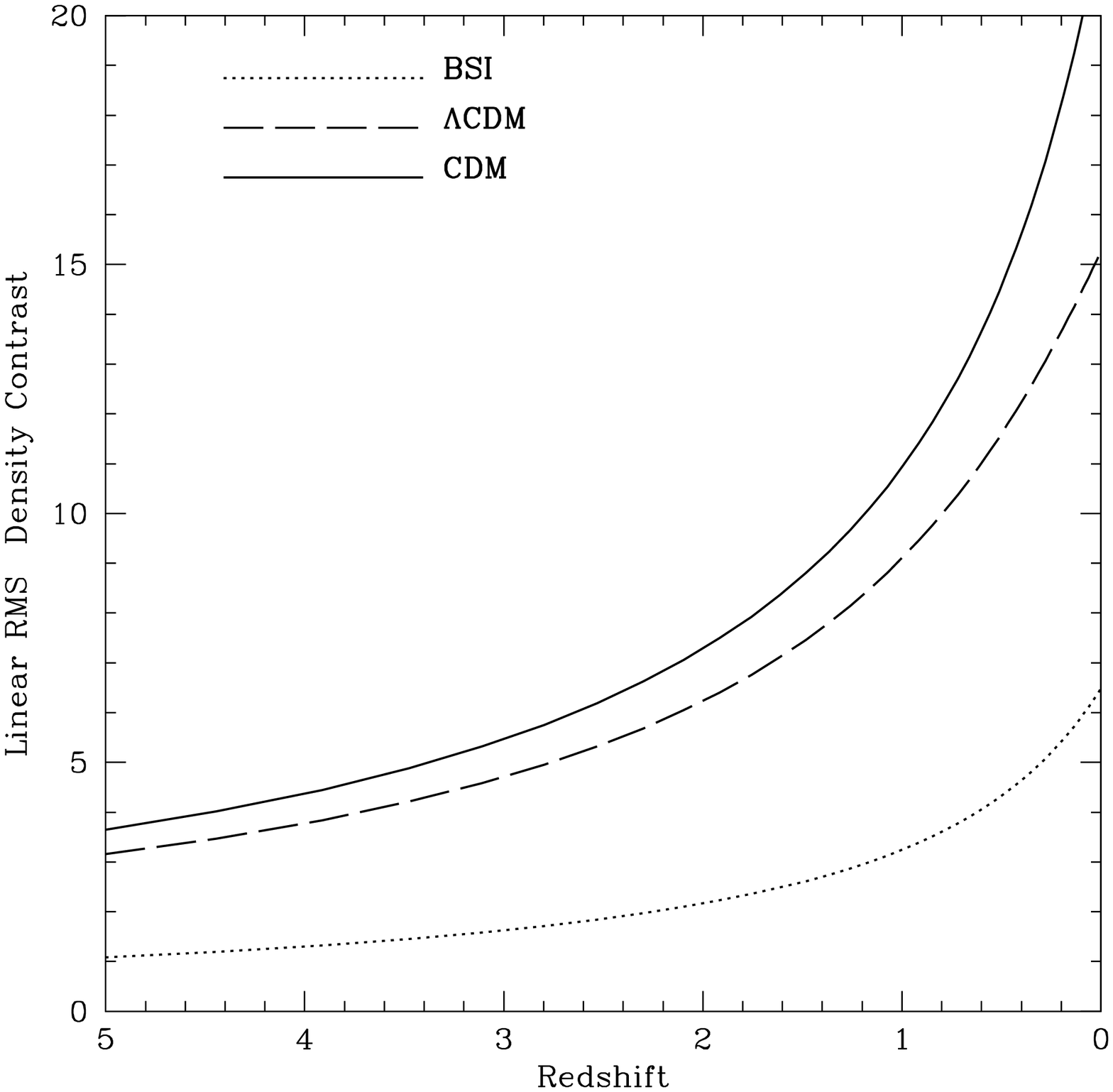,height=\ptamanio}}
\caption{Redshift evolution, of the RMS density fluctuations in a 5 Mpc 
  box as  predicted by linear theory, for the three cosmological scenarios.} 
\label{fig:growthfunc}
\end{figure}

  A partial explanation is provided within the linear perturbation
  theory. In  Fig \ref{fig:growthfunc} we plot the   growth of  the rms 
  density fluctuations, as predicted by linear theory,  in the 3 models. As 
  can be seen,  the  growth rates for both $\Lambda$CDM and CDM are very
  similar up to $z\sim 1$, when the effects of the
  cosmological constant become relatively important 
(e.g. \cite{peebles}).  
Before this epoch,  there are some differences in
the linear growth of fluctuations due to different 
power  spectra and matter content. Nonetheless, similar
structures are  developed in both models up to $z\sim 1$.
 At later epochs, the 
 growth rate of the density contrast  in $\Lambda$CDM decreases  almost to zero ,
 which would explain the difference between its 
collapse time dependence and that of CDM.

\begin{figure}
    \centerline{\psfig{file=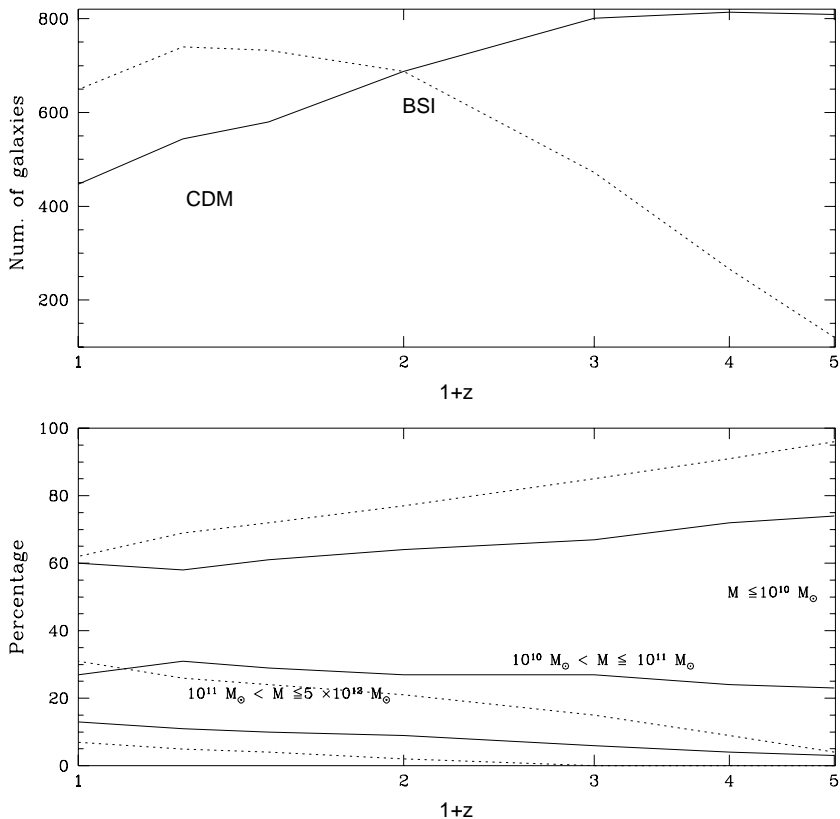,height=\ptamanio}}
    \caption{Redshift evolution of the  total number of {\em numerical}
      galaxies found in all realizations of the BSI and CDM simulations
      (upper panel).  Lower
panel shows the evolution of the  percentages of galactic type halos
for three different mass ranges, in  the BSI (dashed line) and CDM
 (solid line) simulations.} 
    \label{fig:porc}
\end{figure}

The time scale for structure formation is also much longer in BSI. At $z=3$,
small zones are generated in which matter accumulates leading to 
some star formation, although the overall distribution of dark matter is still rather uniform.
It is not until about $z=1$ that a population resembling galaxies can be
identified. 
Fig. (\ref{fig:porc}) shows the number of {\em numerical}  galaxies formed 
in all the BSI and CDM simulations at various redshifts (upper panel).
In CDM, the number of galaxies formed has its maximum at $z=3$. Subsequently,
the coalescence rate exceeds the formation rate, leading to a decrease of the
total number.  In contrast, the number of galaxies in 
BSI does not attain a maximum till 
$z \sim 0.25$, reaching a value similar to the maximum
in CDM.


The galaxy mass function is shown in the lower panel.  In CDM,
the percentage of halos with 
mass in the range 
$10^{10}M_\odot < M \leq 10^{11} M_\odot$  is about 23\% for $z=4$;
already at this 
redshift there are quite massive galaxies present (3\% of the galaxies
with masses exceeding  
$10^{11} M_\odot$). In contrast, the BSI model in this box 
only yields 4\% of galaxies with a mass in the range
 $10^{10}M_\odot < M \leq 10^{11} M_\odot$, 
 and  
not a single one more massive than $10^{11} M_\odot$ at $z=4$.
This relative delay in generating objects poses serious difficulties for this
model, considering the presence of objects detected at $z \sim 4$ in some 
observations \cite{valls}. 

Returning to the similarity in structures generated in BSI at $z=0.25$ with 
those of CDM at $z=3$, we note that the relative delay in evolution is
directly related to the difference in amplitude, $\delta \varepsilon=9$,
between the power spectra $P^{CDM}(k)$ and $P^{BSI}(k)$ 
at the scales of interest (Fig. (\ref{fig:esp2})); in linear theory, one may obtain
an expression relating the epochs of similarly developed structures in
the two models: 
\be
z^{CDM} = \sqrt{\delta \varepsilon} (z^{BSI}+1) - 1.
\ee
Hence, the visual impression of similar structure at different redshifts 
can be explained by the relationship between the spectral amplitudes of the models.

\subsection{Distribution of the baryonic component}
\label{sec:baryoncomp}
\begin{figure}
    \centerline{\psfig{file=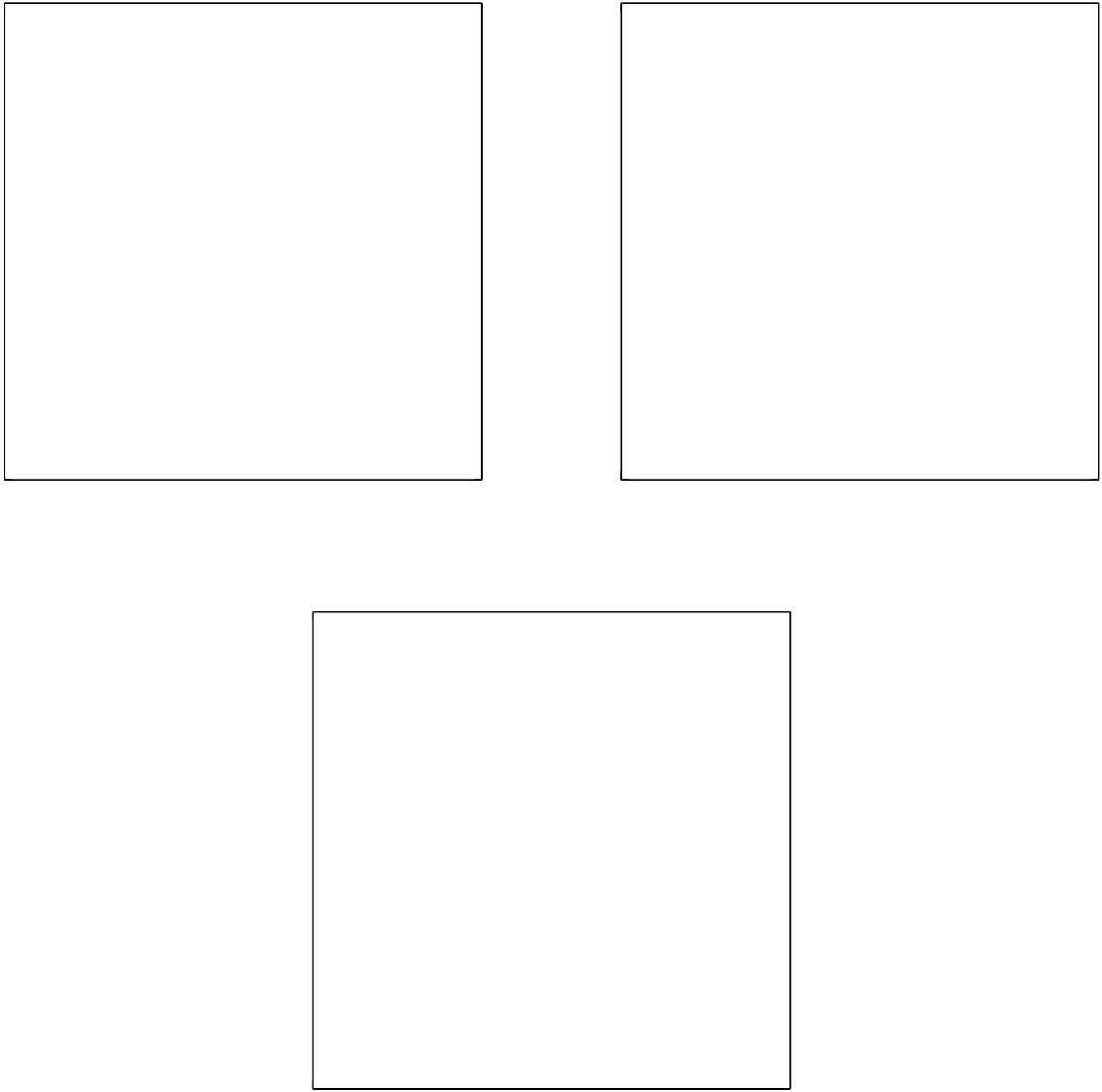,height=\pptamanio}\hspace{5mm}
\psfig{file=EGKK_fig9.ps,height=\pptamanio}} \vspace{0.5cm}
 \centerline{\psfig{file=EGKK_fig9.ps,height=\pptamanio}}
    \caption{3D  Isodensity contours of dark matter (transparent grey) for
  $\rho_{dm} = 30 \;
      \rho_{cr}$ and of star density (yellow shading) 
for $\rho_* = 5 \; \rho_{cr}$ corresponding to the realization 67736
at $z=0$ in the three different scenarios: (a) CDM, (b) $\Lambda$CDM
and (c) BSI. The 
region  shown corresponds to  the position of the most massive halo 
generated in the simulations. 
Full color figures 
 are available in GIFF format at:{\tt
   http://astrosg.ft.uam.es/$\sim$gustavo/newast}} 
    \label{fig:stdm}
\end{figure}

Up to now, the information discussed is of the sort available from
numerous N-body simulations. The inclusion of the various gas components in
our simulations provides
a basis for describing the properties of the baryons. Although the motion of the
starburst particles is simulated by N-body techniques, their
generation is directly related to the dynamics of the 
gas.   
Fig. (\ref{fig:stdm})
 shows
%
%
the densest region of simulation 
67736 at $z=0$ in the three cosmological scenarios. It shows density 
isocontours of dark matter for $\rho_{dm} = 30
\rho_{cr}$ and of star density for $\rho_* = 5 \rho_{cr}$. Note that 
number of halos varies from model to model, being larger in BSI and smaller in
$\Lambda$CDM.   Many halos are not capable of generating enough stars;
these dark halos 
are not seen in the dark matter particle distribution because of the
low percentage of particles 
plotted. BSI exhibits a larger number of halos with stellar formation
as well as 
a more homogeneous distribution and less merging, corresponding to 
a hierarchical structure scenario at a relatively early stage of evolution.



It is interesting to note in Fig. (\ref{fig:stdm}) that
the ability of the numerical model to  generate a stellar component in the halo
provides us with much more complete information about the
internal structure of the halo.  In this way, for the CDM scenario it
is possible to  
identify two components  of star formation in the interior of a halo, thus
allowing one to describe the internal structure in much more 
detail than simply using dark matter. In particular, this type of 
analysis allows identification of
several galaxies within one halo, which is important for 
drawing conclusions regarding the degree of collapse  in a given 
galaxy formation model.  This is an important advantage 
compared to N-body dark matter simulations, in which interacting halos mix and
do not allow precise determination of  the existence of several
components in the same halo.


\begin{figure}
    \centerline{\psfig{file=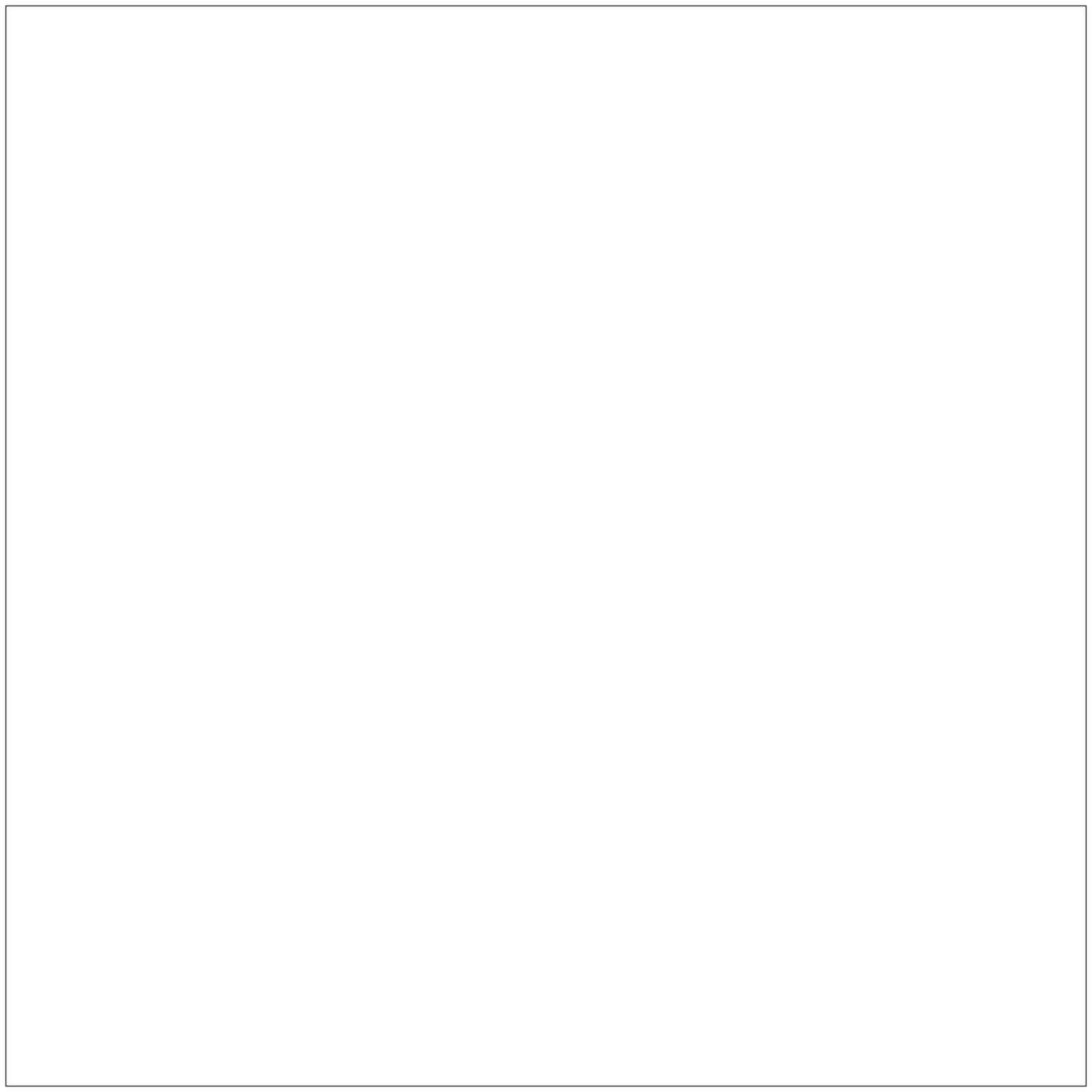,height=\pptamanio}\hspace{5mm}
\psfig{file=EGKK_fig10.ps,height=\pptamanio}} \vspace{0.5cm}
\centerline{\psfig{file=EGKK_fig10.ps,height=\pptamanio}}
    \caption{3D isodensity contour of  the gas distribution for
  $\rho_{\rm gas} = \rho_{cr}$ corresponding to the realization  67736 
at $z=0$ in the three cosmological scenarios: (a) CDM, (b) $\Lambda$CDM
and (c) BSI. 
The surface has been colored  according to the gas
temperature in this isodensity contour: {\em red} $T \sim 5 \times
10^5$ K, {\em yellow} $T \sim 4
- 3\times 10^5$ K, {\em green} $T \sim 2.5 \times 10^5$ K and {\em blue}
$T \sim  5 \times 10^4$ K. Full color figures 
 are available in GIFF format at:{\tt
   http://astrosg.ft.uam.es/$\sim$gustavo/newast}}
    \label{fig:gasovd1}
\end{figure}

\begin{figure}
    \centerline{\psfig{file=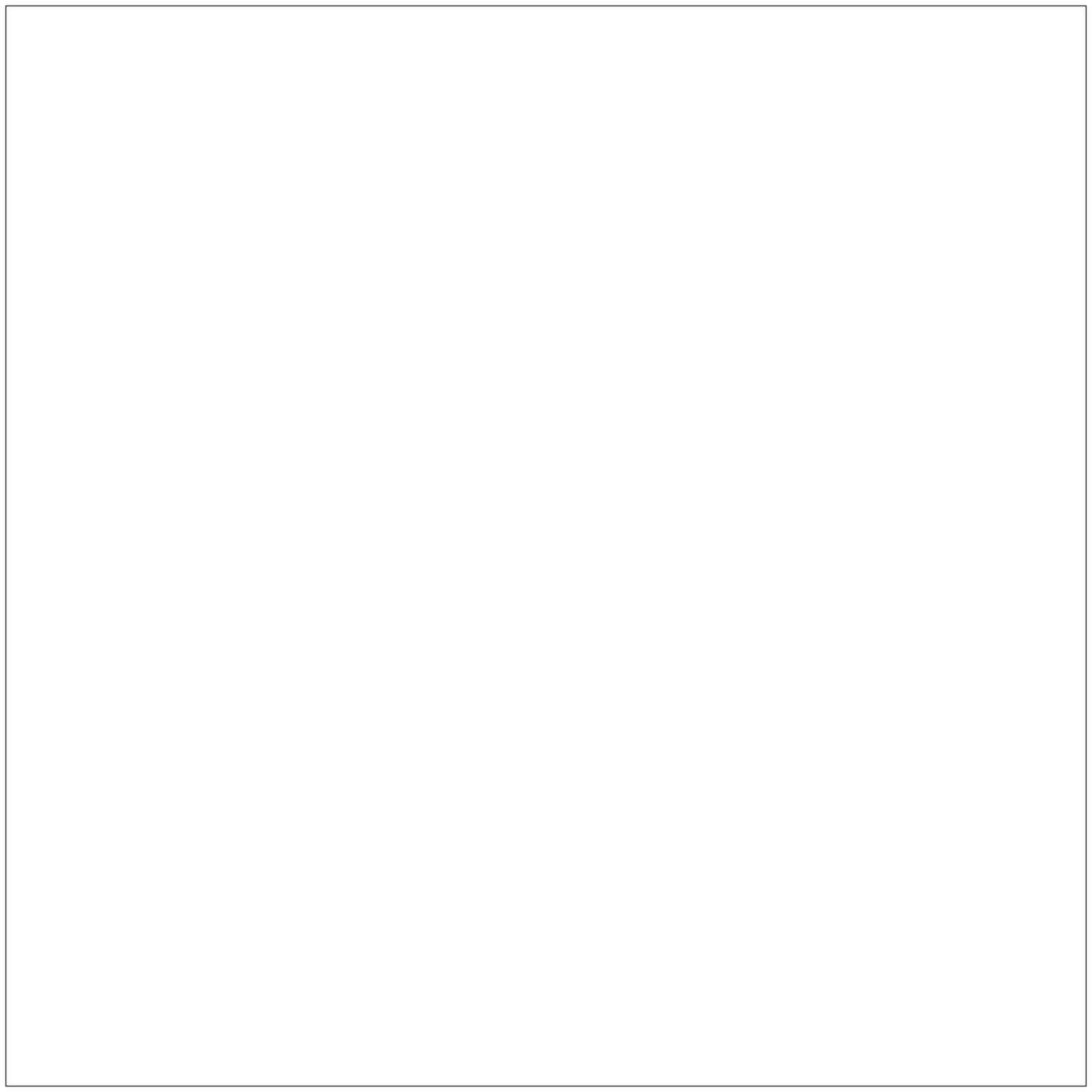,height=\pptamanio}\hspace{5mm}
\psfig{file=EGKK_fig11.ps,height=\pptamanio}}\vspace{0.5cm}
\centerline{\psfig{file=EGKK_fig11.ps,height=\pptamanio}}
    \caption{Same as Fig. \protect\ref{fig:gasovd1} but for
 $\rho_{\rm gas} =  0.5 \; \rho_{cr}$. The  regions of gas temperature
$T \sim 2 \times 10^5$ K are also  shown, (transparent light brown
color surface). Full color figures 
 are available in GIFF format at:{\tt
   http://astrosg.ft.uam.es/$\sim$gustavo/newast} }
    \label{fig:gasovdt}
\end{figure}

A description of the gas distribution and temperature 
provides information on the dynamical evolution of the gas,
as well as on the locations and the conditions under which stars are generated.
Fig. (\ref{fig:gasovd1}) illustrates the $\rho_{\rm gas} =
\rho_{cr}$ density isocontour. To gain more insight into 
gas dynamics, this surface has been 
shaded according to the gas temperature in the isocontour.
The temperature ranges from $ T
\sim 5 \times 10^5$ K (red shading) , $ T \sim 3 - 4 \times
10^5$ K (yellow), $ T \sim 2.5 \times 10^5$ K (green) and $ T \sim 5
\times 10^4$ K
(blue).
An isocontour of lower gas density $\rho_{\rm gas} = 0.5 \rho_{cr}$
is also shown in Fig. (\ref{fig:gasovdt}) with the same
color scheme.   In addition, the regions of gas temperature 
$ T = 2 \times 10^5$ K (cloudy structure shape) are shown. The two figures
reveal a predominant filamentary structure for the low-density gas regions.
These filaments become more evident at lower density contours.


The most evident difference among the three scenarios is the existence of 
more filaments in BSI, compared to a lower number in CDM and 
$\Lambda$CDM, where the filaments are depleted. A general
characteristic of the gas distribution is that the temperature of the density
contour is higher where filaments intersect.  This higher temperature can
be explained as a consequence of accretion shocks (see below) 
and  partly  due to the explosion of 
supernovae produced within halos.  The low-density gas at a temperature
of $T \sim 10^5$ K tends to expand. One sees that this gas expands into 
regions devoid of filaments. Considering that we are comparing the same density 
contours in the three scenarios, we note a significant difference in
the typical volume of faint gas surrounding halos in the respective
scenarios. This volume is largest in CDM as a consequence of
the higher collapse rate, as seen in Fig. (\ref{fig:stdm}) .


\begin{figure}[t]
    \centerline{\psfig{file=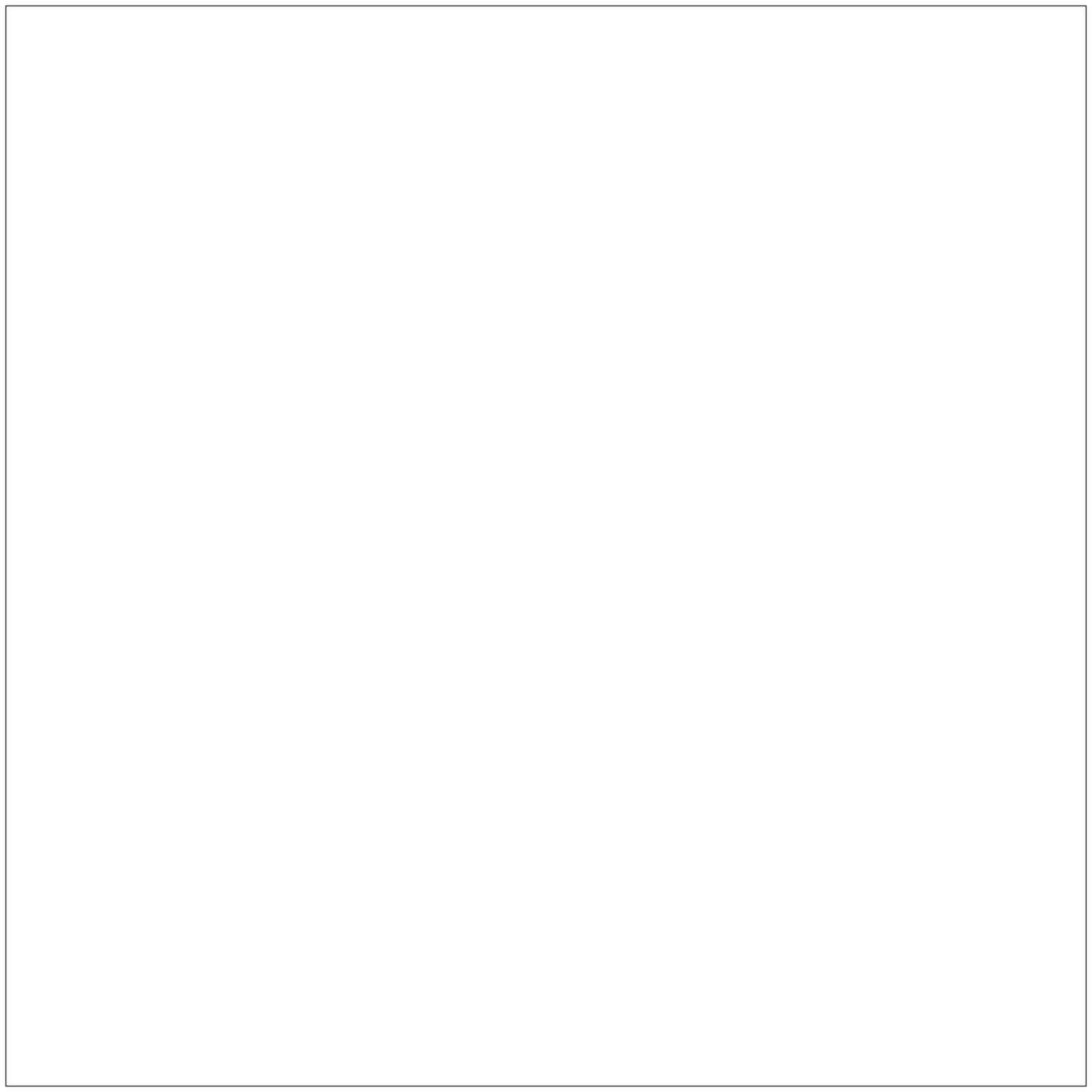,height=\pptamanio}
\hspace{5mm}\psfig{file=EGKK_fig12.ps,height=\pptamanio}}\vspace{0.5cm}
\centerline{
\psfig{file=EGKK_fig12.ps,height=\pptamanio}}
    \caption{ Same as Fig. \protect\ref{fig:gasovd1} but for
  $\rho_{\rm gas} =  0.25 \; \rho_{cr}$.   The region shown is the same
  as in Fig. \protect\ref{fig:stdm}. Full color figures 
 are available in GIFF format at:{\tt
   http://astrosg.ft.uam.es/$\sim$gustavo/newast} } 
    \label{fig:gasovd025}
\end{figure}


\begin{figure}[t]
    \centerline{\psfig{file=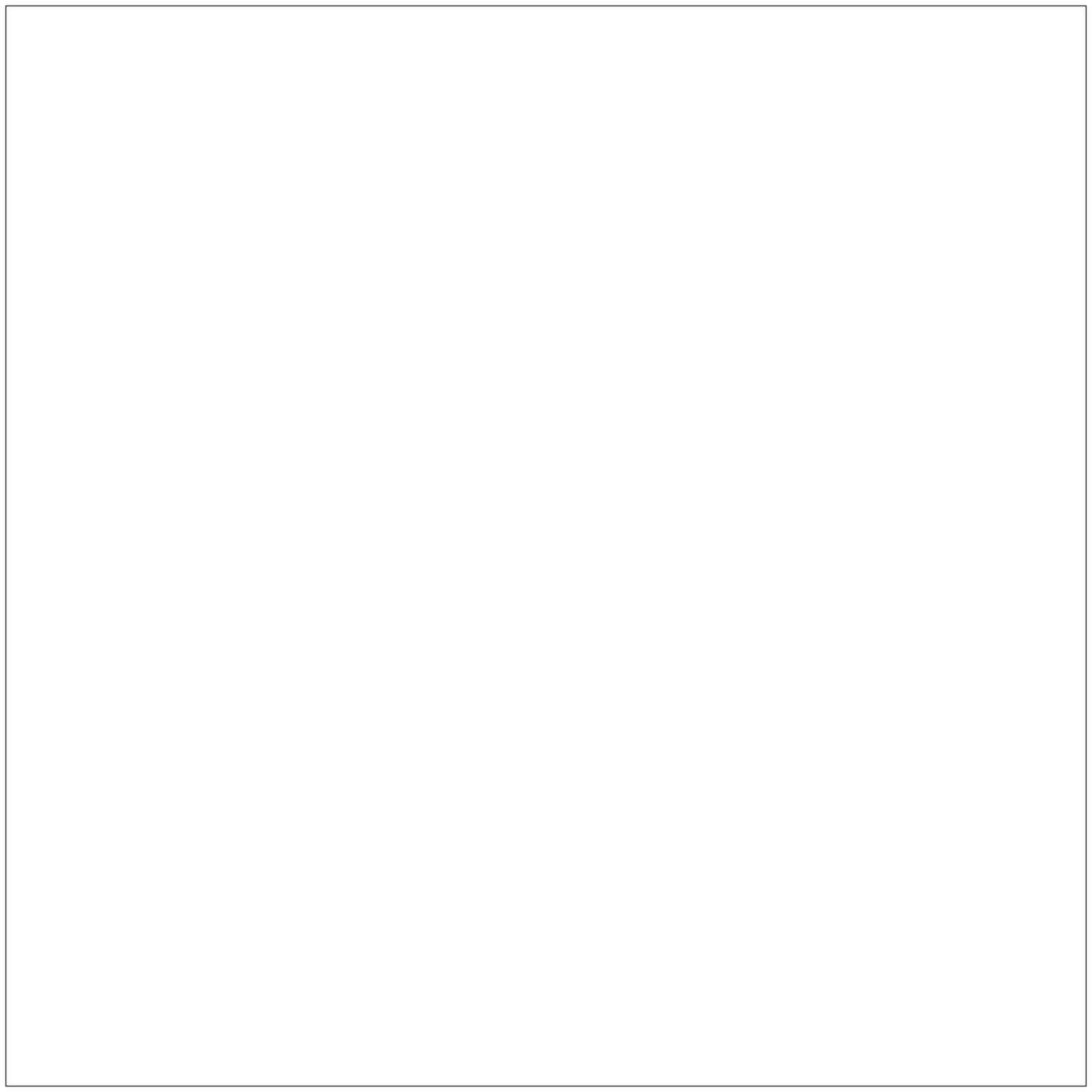,height=\pptamanio}
\hspace{5mm}\psfig{file=EGKK_fig13.ps,height=\pptamanio}}\vspace{0.5cm}
\centerline{
\psfig{file=EGKK_fig13.ps,height=\pptamanio}\hspace{5mm}\psfig{file=EGKK_fig13.ps,height=\pptamanio}}
    \caption{A  slice cut through the most
massive halo of realization  67736 of    CDM   for three
simulations with different  feedback parameter: (a) $A=200$, (b) $A=50$,
(c)  $A=0$ and (d) No feedback.
  Gas temperature  in this slice has been
color coded: {\em white} $T \geq 1.5 \times 10^6$ K, {\em red} $T \sim 1-1.5 \times 10^6$ K, {\em yellow} $T \sim 
7 \times 10^5$ K, {\em green} $T \sim 4 \times 10^5$ K and {\em blue}
$T < 2.5 \times 10^5$ K.  The gas velocity field is also shown.
Arrow sizes are  proportional to
velocity moduli. Black line contours represent the gas density in this
slice and  are equally spaced in steps of  $0.03 \rho_{cr}$. The
position  of the massive halo has been shifted  to the center 
of the simulation box in order to
show the whole structure around it.  }
    \label{fig:vec}
\end{figure}

In Fig. (\ref{fig:gasovd025}) we show a gas density contour of $0.25
\rho_{cr}$. The region shown is as in Fig. (\ref{fig:stdm}), i.e., 
the position of the most massive halo generated. Because of the lower 
value of the density on this contour, one sees that the filamentary structure
is more prominent. Superimposing the two figures demonstrates how 
dark matter halos are located in the  bubbles of
filaments. Moreover, those halos which have generated stars tend to be
found in gas regions of hotter gas, illustrating the relation between 
active star formation and heating of the gas. We may conclude that
star formation predominates in the most massive galaxies as well as in
lower-mass 
objects located in relatively dense filaments. Isolated low-mass halos 
generate fewer stars,
because the supply of  gas is depleted after the first stars form and
is not replenished. The degree of this effect depends of course on 
supernova feedback. For low values of $A$, large pressure gradients
tend to develop, expelling gas more efficiently from  shallow
potential wells.

As can be seen in the figures, the characteristic large-scale structure 
exhibited by the gas distribution is filamentary. However, even if the
density  
contour is lowered somewhat, the dark matter does not exhibit such structures. 
One would need to go to densities lower than the 
mean density to see such structures in the dark matter distribution.
In order to explain why the gas shows this morphology, 
we recall the arguments of Bond   {\em et al.} \cite{bond}, who demonstrate
that the first structure formed due to the growth of dark matter fluctuations
is the filamentary distribution.  This initial structure imposes itself
on the gas by gravitational attraction.  Meanwhile, the dark matter continues to
evolve, falling toward the nodes at the filament intersections
(higher density regions), generating halos and eventually depleting the 
filaments.  However, the gas does not fall freely due
to the opposing force of pressure gradients and tends
to trace the filaments despite their
rather low density contrast in the dark matter component.

An observation with important implications
for this paper concerns how gas flows around  massive halos.
 In Fig. (\ref{fig:vec}),  we show the
temperature,  gas density contours and the gas velocity  field for
a slice cut through the center of the most massive halo found in the
CDM 67736 simulation. 
In order to see the effects of the supernova
feedback loop on gas dynamics, we plot slices corresponding to the
same simulation run with 3 different $A$ values. (In these
figures, the box coordinates have been transformed to 
put the halo at the center.)
The low-density, high-temperature,   gas occupies   
the void regions surrounding the
halo, generated by accretion  shocks which tends to impede 
the penetration of this tenuous gas toward the center.  Most of the gas  
which is incorporated into the halo enters along the filaments
in this case.   This observation does not support  
the hypothesis  of spherical collapse,  which is one of the main assumptions
of semi-analytical models \cite{kauff,lccl} implying at the very least 
that this hypothesis is not always an accurate approximation to the dynamics. 
 
The ``cavern'' around the massive halo,  delineated by the   shock
fronts,    is  bigger ($\sim 1.5 $ Mpc in size) 
 and the gas is hotter for the simulation with  $A=50$, 
which corresponds to high reheating of the gas from supernovae
explosions (see YK$^3$). Thus, the gas temperature at the shock front
is $\sim 10^6$ K for the $A=50$ simulation, while for the $A=0$ and
$A=200$ simulations the temperature drops to $\sim 1-4 \times 10^5$ K
respectively. 
Another feature that can be appreciated in
these figures is that the  gas density gradient is less steep around
the  halo for simulations with supernova feedback than for the
simulation without it ($A=0$).  Note, however that the simulation with
$A=0$ was run  taking  into account the effects of metal
enrichment and enhanced cooling,  as  described in Section
\ref{sec:hydromodel}. In order to check  the effects of our modeling
of   metal production, we have also rerun the same simulation
assuming primordial composition everywhere, as well as with no feedback. In Fig
\ref{fig:vec}-(d),  we plot the results for this simulation. Two
important effects are clearly visible. On the one hand, the size  of the 
cavern  is considerably larger than for $A=0$ with metal
enrichment (Fig \ref{fig:vec}-(c)), and the shape is more
spherical. The temperature of the gas at the shock front is
comparable to the simulation with high reheating from supernovae  (Fig
\ref{fig:vec}-(b)). The gas has expanded in this simulation with
respect to the $A=0$ simulation due to the larger pressure
gradients.  Star formation is considerably lower in the central halo,
when primordial composition is assumed.  Luminosity for this halo is
$\sim 1.5$ magnitudes fainter than for the corresponding halo in the 
simulation with $A=0$ and metal enrichment. 
These results show that effects of metallicity enrichment are important
and cannot be neglected in simulations with star formation.

\section{Effects of environment on galaxy properties}
\label{sec:env}

\subsection{Characterization of galaxy environment}

There are many aspects of ''environment" that one might wish to characterize
and study.  Here, we consider perhaps the most basic characteristic 
of environment, which is straightforward to quantify in an objective way:
the average density of dark matter in the
region about the halo (referred to as ''ambient density" in what follows).  
Specifically, we convolve the density on the grid 
(with origin at the center of mass of the halo) with
a Gaussian window function of characteristic width $\sigma= 0.5$ Mpc:
\begin{equation}
\langle \rho \rangle = \frac{\int \rho \; W(r) \; d^3{\bf r}}{\int W(r)
\; d^3{\bf r}} ; 
\hspace {1cm} {\rm where} \hspace{0.5cm} \; W(r) = \exp (-r^2/2 \sigma)
\end{equation}

\begin{figure}
    \centerline{\psfig{file=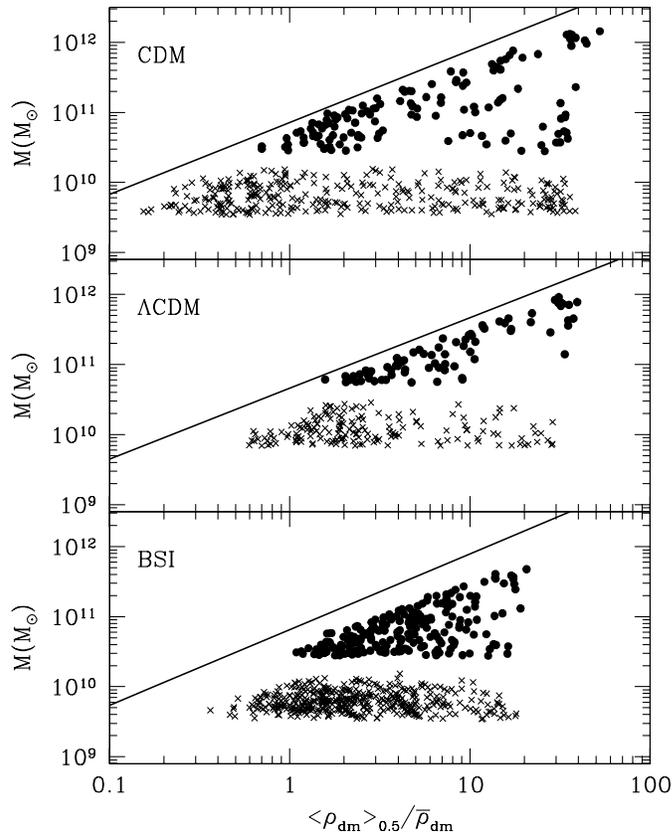,height=\ptamanio}}
    \caption{Relation between halo mass and ambient dark matter
overdensity. The ambient density $(\langle      \rho \rangle)$ is
computed by convolving the grid density with a Gaussian filter of width
0.5 Mpc.      The overdensity is the ratio of $(\langle      \rho
\rangle)$ to  $(\bar{\rho}_{dm})$, the mean dark matter density in the
simulation box . Points represent halos computed in a two-cell radius;
crosses denote halos in a one-cell radius, according to our galaxy
finding algorithm (see text). Solid lines  represent  the minimum
overdensity  a halo would have if it were completely isolated.}
    \label{fig:masaenvmo}
\end{figure}

In Fig. (\ref{fig:masaenvmo})  we plot halo mass vs. ambient  
dark matter overdensity for all the halos found in our simulations.
  Less massive
halos (1-cell radius)  are distributed through the full range of densities.
The solid lines in the figure represent the minimum dark mass overdensity 
due simply to the presence of a halo 
of a given mass, i.e., if it were completely isolated (no dark matter
concentration outside the halo volume within a distance of 0.5 Mpc).
For a halo of a given  mass,
the farther a point is separated from this limiting line, the higher the
ambient density. 
For the less evolved BSI model,
most halos are located in filaments (see  {\S}
\ref{sec:darkdistr}), whose dark matter distribution is not very clumpy. 
Therefore, the overdensities 
of BSI halos on a 0.5 Mpc scale are almost always larger than those of 
a corresponding isolated object. For CDM, in contrast,
the dark matter distribution is clumpier, and a larger proportion of 
halos are located in rather isolated regions.



The dark matter densities in which halos are found range from $0.1
\bar{\rho}_{dm}$  
up to about $\ 50 \bar{\rho}_{dm}$ in CDM and $\Lambda$CDM. 
This range of densities corresponds
to the environment of field galaxies up to small groups at the high end.
The dark matter distribution at $z=0$ in BSI is more homogeneous than in 
the other two scenarios due to the less 
advanced stage of evolution, and the range of surrounding densities is
correspondingly lower.

\subsection{Dependence of  morphology of galaxies on environment}
\label{sec:mfenv}

The relationship between the properties of galaxies and their
environment is a controversial subject that has been studied from a
number of theoretical aspects.  
These include morphology - density (MD) relations, i.e., correlations 
between the probability of finding 
galaxies of particular morphological types and
the density of their local surroundings.
\cite{morfd}. The dependence of the fraction of galaxies belonging to the
various morphological types on their environment is well established
observationally. This dependence is especially evident 
when one compares populations of field and cluster galaxies 
\cite{depenvc} or of clusters of varying richness
\cite{oem,morfd}. Observations have established that, both inside and 
outside of clusters, the abundance of elliptical and lenticular galaxies
relative to spirals increases as a function of the density of the
environment.   

Of particular interest in the context of cosmology and
large-scale structure are those effects resulting from 
interactions between galaxies, encompassing phenomena ranging from 
small perturbations due to relatively distant neighbors
up to strong distortions caused by close encounters, galactic cannibalism,
or merging of systems of comparable mass.
Examples of such processes are frequently seen in the simulations
reported here (see Figs.(\ref{fig:3ddark}) and (\ref{fig:3dstars})).   

A review of processes that could contribute significantly to MD relations
may be found in \cite{pegarse}. These include the following:
\begin{itemize}

\item Consequences of the initial conditions of formation, such as 
more rapid star formation during the collapse of galaxies falling
toward the center of a protocluster, leading to elliptical galaxies
in the center.

\item Consequences of subsequent evolution.

\item Interactions with the environment,
such as 1) compression and evaporation of gas in galaxies passing through dense regions,
thus inhibiting star formation and producing lenticular galaxies at the expense of spirals, 
or 2)  loss of gas and stars due to tidal interactions in galaxies falling
toward the center of a cluster and finally resulting in formation
of galaxies of type cD.

\item Interactions between neighboring galaxies, including 1)
merging of spiral galaxies resulting in the formation of ellipticals at
the center of a cluster or 2) galactic cannibalism of cD galaxies.

\end{itemize}

Observational evidence for the formation of elliptical galaxies due to
merging of spiral  galaxies has been accumulating during the last 10 years
 \cite{fusionez}.

The existence of interactions is firmly established both
observationally and theoretically.  Nonetheless, certain questions  
crucial to proper modeling of galaxies
have not yet been resolved satisfactorily, such as 1) the frequency of such
events both at present and in the past and 2) whether these
interactions 
are an essential or even dominant feature in galactic evolution.
In other words: To what extent are the observational properties 
of galaxies governed by environmental effects as opposed to initial 
conditions? 

Hydrodynamical simulations can clearly contribute to our theoretical understanding 
of MD relationships: For the range of densities covered by our simulations, the main
mechanisms influencing the MD relationship are mergers, tidal forces, and other interactions.  
A typical illustration may be found in Fig. (\ref{fig:stdm}).

Our procedure is as follows: 
Due to the limited spatial resolution of our study, which prevents a direct
morphological classification, we have  assigned a morphological type to
each halo identified in the  
simulations according to the position it occupies in the UBV color diagram.
(Of course, one should be aware that due to the large dispersion in the 
range of colors corresponding to each morphological type, this assignment
on the basis of color gives only an approximation to the true morphology.)
Once we have a precise working definition of the "environment" of the halos and their "morphological
classifications", it is then possible to analyze 
the  morphological dependence of galaxies on environment.


\begin{figure}
    \centerline{\psfig{file=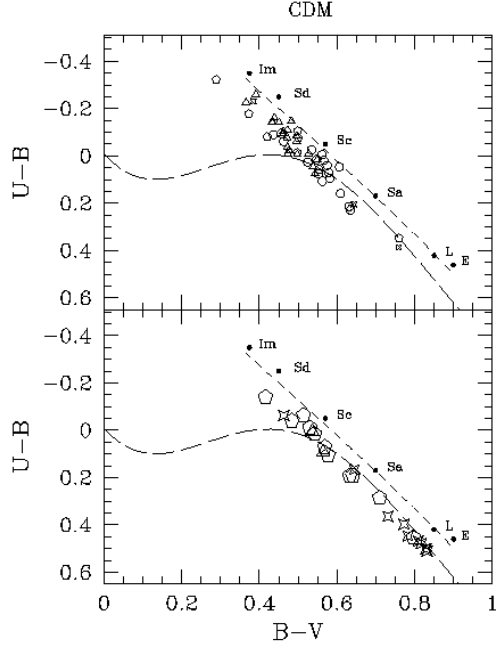,height=\ptamanio}}
    \caption{Color-color diagram for galaxies brighter than       $M_B
\leq -16$ in the CDM simulations. The long-dashed line       represents
the UBV position for main-sequence stars of luminosity class V.        The
short-dashed line provides a reference for galaxies. Along this line,
galaxies of different morphological types are indicated in the position
corresponding to their average colors. Upper panel: galaxies with $M
\leq  3 \times 10^{11} M_{\odot}$; lower panel: galaxies with       $M
>  3 \times 10^{11} M_{\odot}$. Galaxies located in environments of
ambient density $\langle \rho      \rangle  /\bar{\rho}_{dm} > 30$ are
denoted by stars; those with      $ 10 < \langle \rho \rangle
/\bar{\rho}_{dm}      \leq 30$ are indicated by pentagons, those with $
3 < \langle \rho \rangle /\bar{\rho}_{dm} \leq 10$ by triangles, and
those with $\langle \rho \rangle /\bar{\rho}_{dm}\leq 3$ by circles.}  
    \label{fig:ccdm}
\end{figure}


\begin{figure}
    \centerline{\psfig{file=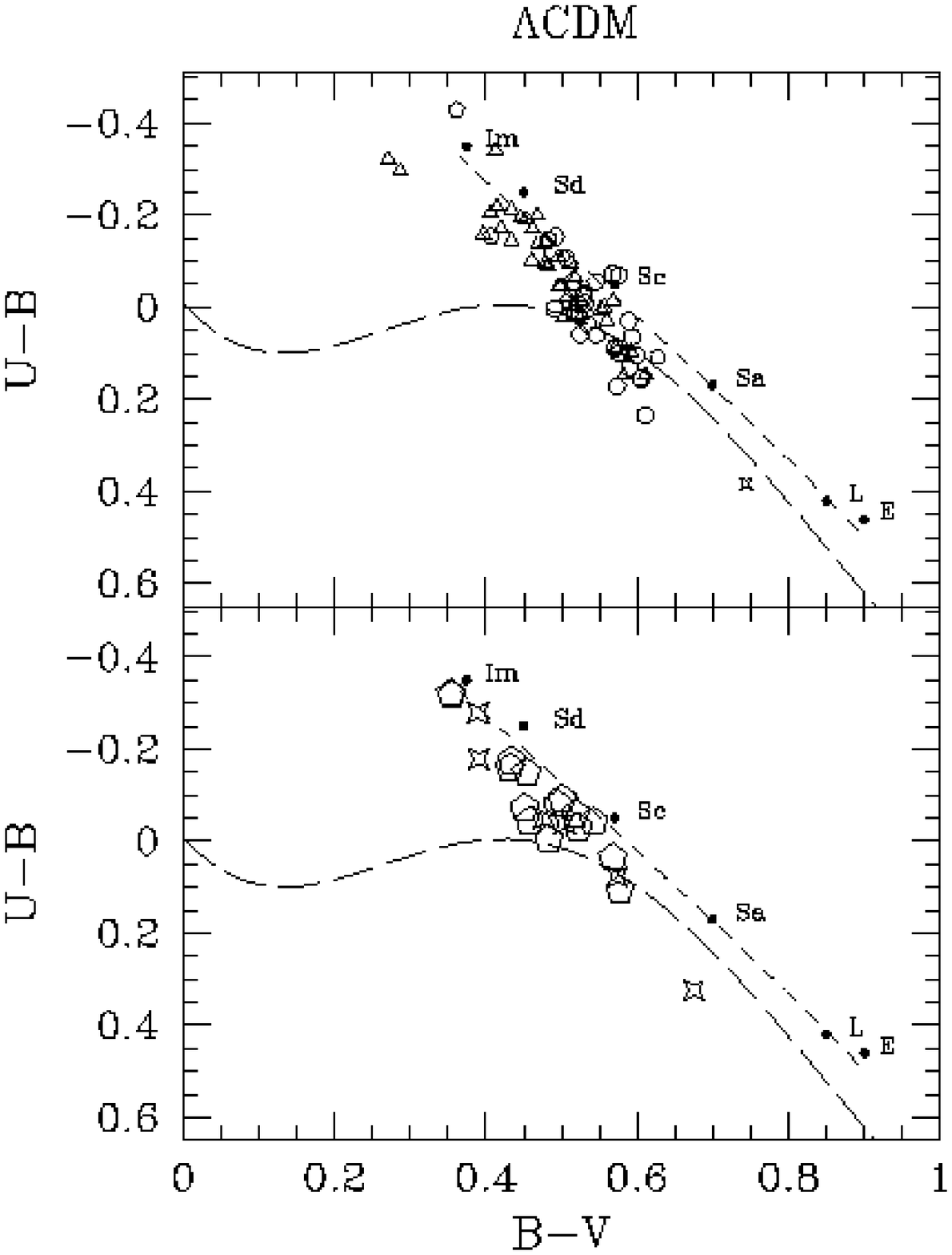,height=\ptamanio}}
    \caption{Color-color diagram for galaxies brighter than       $M_B
\leq -16$ in the $\Lambda$CDM simulations. Panels and symbols as in Fig.
\protect\ref{fig:ccdm}.}     
\label{fig:clcdm}
\end{figure}


\begin{figure}
    \centerline{\psfig{file=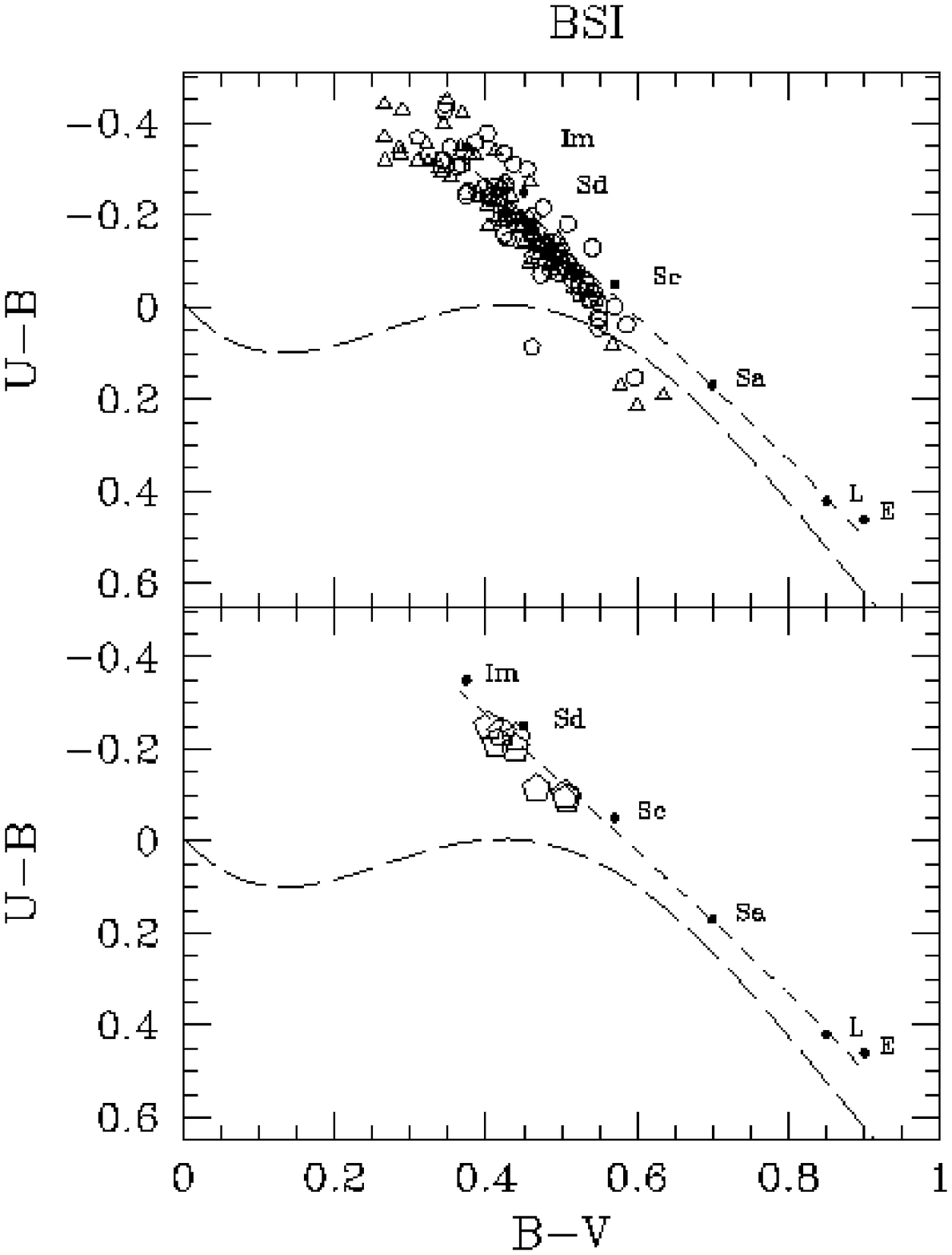,height=\ptamanio}}
    \caption{Color-color diagram for galaxies brighter than  $M_B \leq
-16$ in the BSI simulations. Panels and symbols as in Fig.
\protect\ref{fig:ccdm}.} 
    \label{fig:cbsi}
\end{figure}

Figs. (\ref{fig:ccdm}),   (\ref{fig:clcdm}) and  (\ref{fig:cbsi}) are
color-color  
diagrams ($U-B \; vs. \; B-V$) for galactic halos compiled from all the
simulations in the respective scenarios. 
Only galaxies with $M_B \leq -16$ are plotted.  
The upper panels show galaxies of mass $M  \leq 3  \times 10^{11}
M_{\odot}$, represented by different symbols according to the density
of their environment. 
The lower panels show galaxies of mass $M >  3 \times 10^{11} M_{\odot}$. 
The long-dashed (lower) curve represents the main sequence for stars of
luminosity 
class V.
The short-dashed line represents colors of galaxies with different 
Hubble morphological sequence index, $T$, originally  introduced  by
de Vaucouleurs \& de Vaucoleurs  \cite{devac} in the {\em 
First Reference Catalogue of Bright Galaxies (RC1)}.
 It ranges    from $+10$ (irregulars) at the upper
left to $-6$ (E0) at the lower right.
Along this line, galaxies of different morphological types
 are indicated in the position corresponding to their average
 colors computed from the RC3 catalogue \cite{devacetal}. The
 dispersion  of color indices  for galaxies of different morphological 
 type ranges from 0.04, for ellipticals  ($T=-5$) up to 0.1, for the
 irregulars ($T=10$). (See Buta {\em et al} \cite{butal} for further details). 

In the CDM scenario (Fig. (\ref{fig:ccdm})) we obtain a correlation between
ambient density and morphology.  For the more massive objects, 
(lower panel) lenticular and elliptical galaxies tend to be generated 
in environments of ambient density $\langle \rho \rangle _{50} > 30
\bar{\rho}_{dm}$.  These objects are relatively isolated from other
bright halos, having 
formed from coalescencing smaller halos. The relationships as
simulated dynamically are thus
consistent with the picture of generation of elliptical galaxies from  
spirals. 

The rest of the halos are distributed along the line
associated with various  
morphologies: in regions of elevated density  ($10 < \langle \rho
\rangle _{50}/ \bar{\rho}_{dm} \leq 30$; pentagons in the lower panel)
one 
finds mostly massive spirals and hardly any irregulars. In the upper panel,
less massive galaxies predominate in low-density environments 
($ \langle \rho \rangle _{50}/
\bar{\rho}_{dm} \leq 3$), characterized by spiral morphology. 
There are also low-mass irregulars, but note that these are 
often located in intermediate or high-density environments ($10 <
\langle \rho \rangle _{50}/ \bar{\rho}_{dm} \leq 30$). 

The $\Lambda$CDM scenario (Fig. (\ref{fig:clcdm})) does not 
lead to lenticular or elliptical galaxies within the range of ambient densities
obtained here (corresponding to groups of galaxies).
This is not necessarily a shortcoming of the scenario, but could simply be a
consequence of our limited dynamical simulation range.
However, in contrast 
to CDM, it does yield massive irregular galaxies, while spirals predominate.
The low-mass halos are also located in low-density environments, with spirals 
predominating.  The absence of ellipticals and spirals of 
type $Sa$ may be attributed to 
the lower merger rate of halos in $\Lambda$CDM compared to CDM.
This fact implies that the probability of finding field ellipticals in
$\Lambda$CDM is much lower than in CDM. In order to be able to  
simulate ellipticals in $\Lambda$CDM, one
would need to model larger volumes containing some regions of higher
ambient density. 

Another peculiarity of $\Lambda$CDM is a greater tendency for blue galaxies, 
contrary to what one might initially expect from a scenario with cosmological 
constant, if galaxies have had more time to evolve and thus become redder.  
Recall however, that here the parameters of each cosmological scenario 
have been chosen to yield an age of the universe similar to the CDM case
 ($\sim 13 \times 10^9$ years) and a comparable time for evolution
of galaxies. We will explain the  relative redness of the CDM galaxy
sample in section \ref{sec:sfrmf}.

Finally, the BSI scenario Fig. (\ref{fig:cbsi}) gives galaxies that are
more concentrated 
in the blue regime. There is an excess of irregulars.  Now, in the box
simulated here, halo environments denser than $20 \bar{\rho}_{dm}$
(Fig.  
(\ref{fig:masaenvmo})) are not realized in BSI as a result of the lower
spectral amplitude. 
(Such environments could be realized in BSI if much larger scales were
simulated.) 
As a consequence, ellipticals are not formed in simulations on this scale.  

Earlier, we remarked that BSI as realized in a small box resembles an
earlier stage  
of CDM.  One consequence is that spiral galaxies of BSI located in
regions of intermediate ambient density (pentagons in the lower panel)
would presumably evolve further and end up as ellipticals and
lenticulars, thus resembling the presently found CDM
structures. Another way of looking at this point is that in a true BSI
universe there would be some dense regions due to nonlinear effects on
much larger scales, and in these regions the evolution could proceed on
a timescale comparable to that of the CDM models realized here.  Hence,
it  is reasonable to suppose that a true BSI universe would indeed produce
ellipticals in dense regions at present.   

In spite of these trends, the theoretical MD relationships emerging from our simulations
are not as simple as might
be inferred from the above remarks.  For example, returning to
Fig. (\ref{fig:ccdm}) 
one can appreciate the quite diverse morphologies produced 
according to the CDM scenario in high-density environments, 
ranging from $Sd$ spirals to ellipticals. Thus,  in addition to reproducing 
known trends quite naturally,
our model appears capable of generating the complexity of conditions required for producing 
galaxies with a range of morphological characteristics. 
Because of the close relationship between morphology
and star formation, we will now look more closely at the dependence of star formation 
on the environment, in order to understand the MD relationship in more detail. 

\subsection{Dependence of the star formation rate on environment}
\label{sec:sfrmf}

The influence of the ambient density 
on the star formation rate (SFR) is not yet 
well understood. At least two effects influencing the SFR in high-density
regions are conceivable 
\cite{marcio}.
On the one hand, the SFR could be enhanced by tidal interactions
triggering star formation, 
possibly in the form of bursts  \cite{bus,ken}. On the other hand, in
the very high density regions 
in the core of clusters, close galaxy encounters lead to a depletion
of interstellar gas and thus preferentially leave
anemic spirals in clusters of galaxies \cite{dress}.

The range of densities obtained for our {\em numerical} galaxies corresponds
to the environment of field galaxies up to small groups at the high end.
In addition, as mentioned earlier, close encounters, galactic cannibalism,
and merging of systems of comparable mass are frequently encountered
processes in the simulations reported here 
(see Figs.(\ref{fig:3ddark}) and (\ref{fig:3dstars})).  Hence,
our study allows analysis of the SFR in an environmental situation
of low density and frequent galaxy encounters.

\begin{figure}
    \centerline{\psfig{file=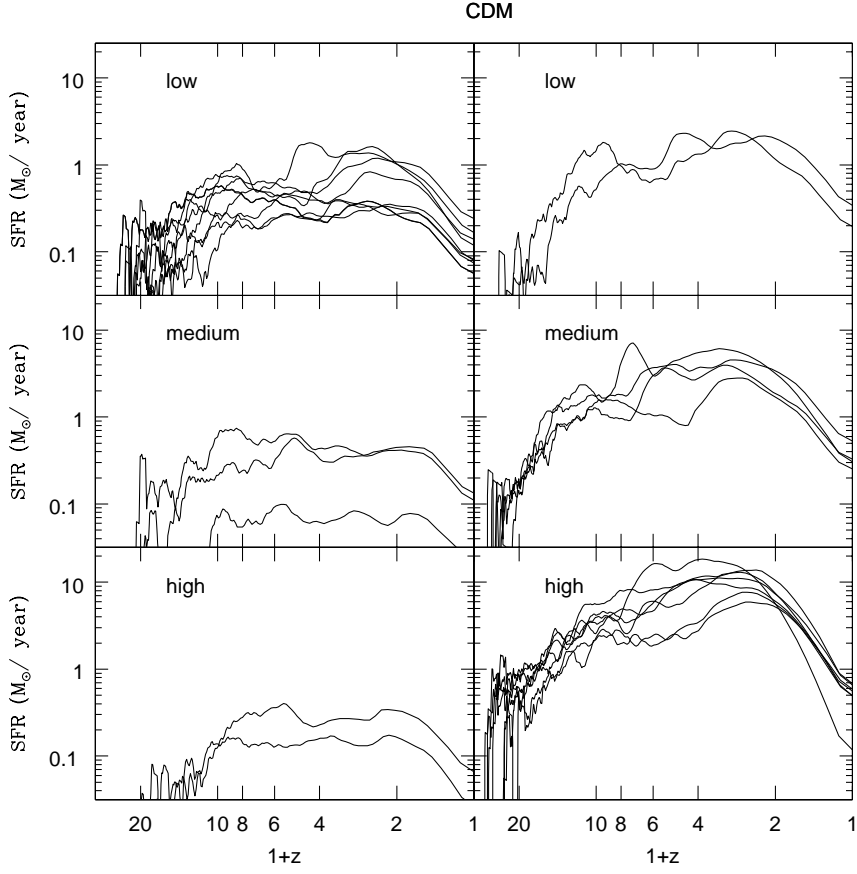,height=\ptamanio}}
   \caption{Redshift evolution of the Star Formation Rate  for  halos
     found in   all  CDM simulations. 
 Left column: galaxies with $M  \leq 3  \times 10^{11}
M_{\odot}$. Right column: galaxies with $M >  3 \times
10^{11}M_{\odot}$. Each column is divided depending upon  the 
density of the environments where galaxies are located: {\it low} density 
($ \langle \rho \rangle _{50}/ \bar{\rho}_{dm} \leq 10$), {\em medium} density
($10 < \langle \rho \rangle _{50}/ \bar{\rho}_{dm} \leq 30$) and {\em
high} density 
($\langle \rho \rangle _{50} /\bar{\rho}_{dm} > 30$). Only galaxies with $M_B
\leq -16$ are plotted.}
    \label{fig:tasacdm}
\end{figure}


\begin{figure}
    \centerline{\psfig{file=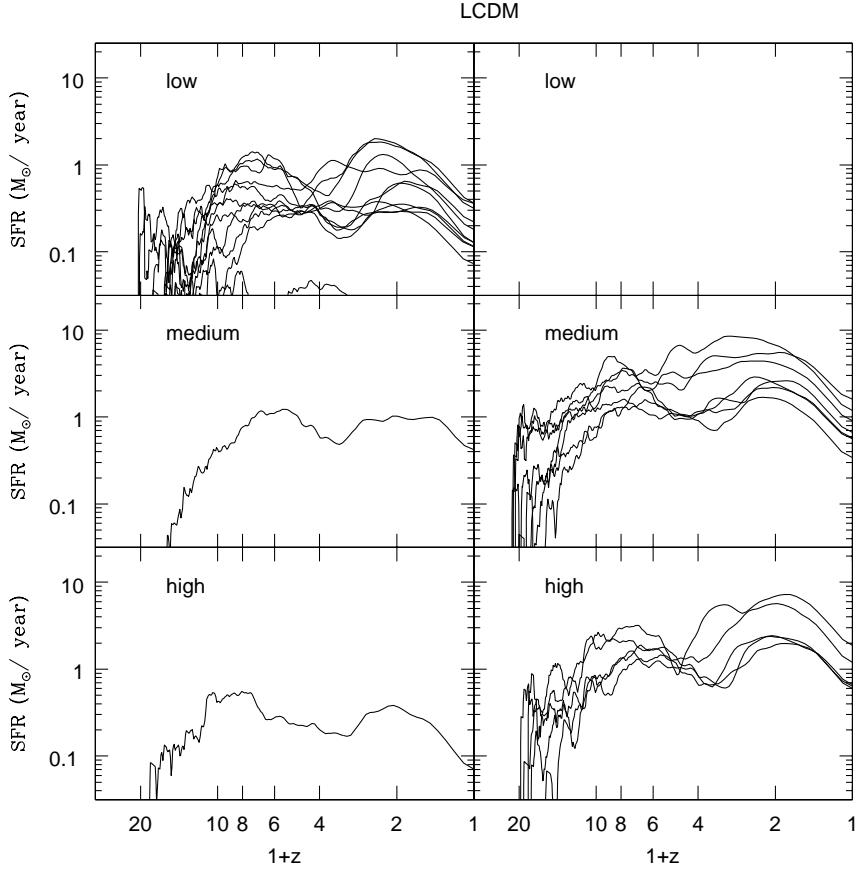,height=\ptamanio}}
\caption{Same as  in Fig. \protect\ref{fig:tasacdm} but for the
$\Lambda$CDM simulations.}
    \label{fig:tasalcdm}
\end{figure}

In Figures (\ref{fig:tasacdm}) and (\ref{fig:tasalcdm}) , the evolution
with redshift of the
SFR for  bright ($M_B \leq -16$) 
 halos found in all simulations for the CDM and
 $\Lambda$CDM models is presented.  The graphics in the left hand
column correspond to galaxies with masses $M \leq 3 \times 10^{11} M_{\odot}$,
while those with $M > 3 \times 10^{11}M_{\odot}$ are shown in the right hand
column.  These groups of galaxies have been subdivided in turn into three
groups according to the ambient density  at their location:  in environments of
{\it low} density ($ \langle \rho \rangle _{50}/ \bar{\rho}_{dm} \leq 10$),
{\em medium} density ($10 < \langle \rho \rangle _{50}/ \bar{\rho}_{dm} \leq
30$) and {\em high} density $\langle \rho \rangle _{50} /\bar{\rho}_{dm} >
30$, (upper, middle, and lower panels, respectively).

Note that this classification according to the 
environment takes into account the value of the density of material
surrounding the galaxy at the current epoch (z=0), whereas the true
density at the redshifts for which the SFR is plotted is not available.
Thus, the current density is used as a surrogate for the density at
the time of star formation and obviously is not a perfect indicator.

As shown in Figures (\ref{fig:tasacdm}) and (\ref{fig:tasalcdm}),
the SFR's of the least massive halos do not exhibit
any clearly defined correlation with their environment.
A plausible explanation is that
the SFR of less-massive halos depends mainly on their capacity
to hold on to the gas, since there are several competing
processes that can inhibit or stimulate star formation.
Since these various processes have comparable magnitudes,
it is not surprising that there are no clear trends in the dependence
of the SFR on environment in less massive halos.

In contrast, for  massive halos 
($M >  3 \times 10^{11}M_{\odot}$ ) in   CDM  simulations, 
a drop of the SFR from  $z \approx 1$ to
$z=0$ is clearly seen. This drop is steeper for halos located in higher-density
environments. A drop is also seen  in $\Lambda$CDM simulations, 
albeit less pronounced. 
Typically, this drop in the star formation rate was associated with
 an apparent depletion of the cold gas available for star formation in the
central regions of the halos. The dynamics of the 
supply of cold gas are complicated due to the supernovae feedback
loop: Cold gas can be depleted if cooling is too slow
compared with heating processes.  Mergers tend
to supply additional heat to the gas in the form of shocks. 
Hence one explanation for 
the star formation drop could be the increased frequency of  mergers,
which are characteristic for this period of evolution (see
Fig. (\ref{fig:3devoldm})). As we have seen
in the course of this paper, in the CDM model
the collapse rate is larger than in 
$\Lambda$CDM, which explains the difference
in the degree of  decrease in the SFR among the models
and the larger number of
reddened galaxies predicted by the CDM model, as we saw in 
Section \ref{sec:mfenv}.

\subsection{Possible dependence of the Tully-Fisher relation on
environment}
\label{sec:tf}

The empirical relationship between luminosity and line width of spiral  galaxies
 \cite{tfart1,tfart2,tfart3} is
one of the most useful tools in cosmology, most notably
when applied to modeling of large-scale velocity
fields and to determination of the Hubble constant
(see, e.g., Strauss \&  Willick \cite{tfrel}). 

There is some evidence supporting the existence
of a "universal" TF relation for field and cluster galaxies.
Bothun {\em et al} \cite{bot}; Richter and Huchtmeier \cite{rich}; 
Giuricin {\em et al} \cite{giu} and Biviano {\em et al} \cite{biv} 
did not observe any dependence of the
exponent of the TF relation in environments ranging from
cluster to field galaxies. However,  it is difficult to exclude the possibility of various
systematic biases such as hidden dependence on environment
\cite{tfart2,tfart3}.

\begin{table}
\caption{Parameters of the linear fits ($a$, $b$) and correlation
coefficients ($r$)  for the residuals of the
Magnitude-Circular velocity relation in different luminosity bands
 as a function of the color
indices of the {\it numerical} galaxies. \label{table2}}
\begin{tabular}{rccccccccccccc}
   \hline  \hline
\small
 &  &  & \multicolumn{3}{c} {\it a} & & \multicolumn{3}{c} {\it b} & &
\multicolumn{3}{c}  {\it r} \\
   \cline{4-6} \cline{8-10} \cline{12-14}
Model & CI & & B & R & I &  & B & R & I&  & B & R & I \\
\hline

    & $U - B$ & & 1.36 & 0.76 & 0.64 & & -0.08 & -0.05 & -0.04 & &  0.58
& 0.42
& 0.38 \\
CDM \\
    & $B - I$ & & 1.34 & 0.77 & 0.66 & & -2.35 & -1.34 & -1.15 & & 0.64
&
0.47 & 0.42 \\ \\
     & $U - B$ & & 1.98 & 1.25 & 1.01 & & 0.11 & 0.05 & 0.05 & & 0.68 &
0.54 & 0.50 \\
$\Lambda$CDM  \\
     & $B - I$ & & 2.17 & 1.41 & 1.24 & &  -3.56 & -2.32 & -2.05 & &
0.76 &
0.62 & 0.58 \\ \\
     & $U - B$ & & 1.63 & 0.83 & 0.65 & & 0.30 & 0.15 & 0.11 & & 0.61 &
0.40 & 0.34 \\
BSI\\
          & $B - I$ & & 1.89 & 1.08 & 0.89 & & -2.89 & -1.65 & -1.36 & &
0.79 &
0.59 & 0.52 \\
   \hline \hline

\end{tabular}
\end{table}

\begin{figure}
    \centerline{\psfig{file=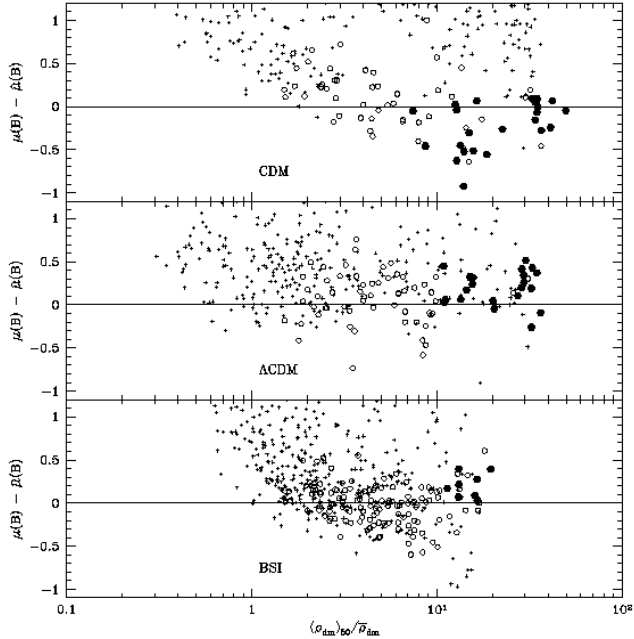,height=\ptamanio}}
   \caption{Residuals, after correcting by color effect (see text), 
 of the magnitude-circular velocity relation in the B band  as a
 function of the  ambient overdensity  $<\rho_{dm}>_{0.5}/\bar{\rho}_{dm}$.
 Galaxies with $M \geq 3 \times 10^{11} M_{\odot}$ are denoted with
dark hexagons,  
open circles represent
halos with $3 \times 10^{10} M_{\odot} < M < 3 \times 10^{11}
M_{\odot}$ and crosses indicate faint halos with $M_B > -16$ y $M \leq 3
\times 10^{10} M_{\odot} $ (not taken into account in the fits).}
    \label{fig:tfresb}
\end{figure}


\begin{figure}
    \centerline{\psfig{file=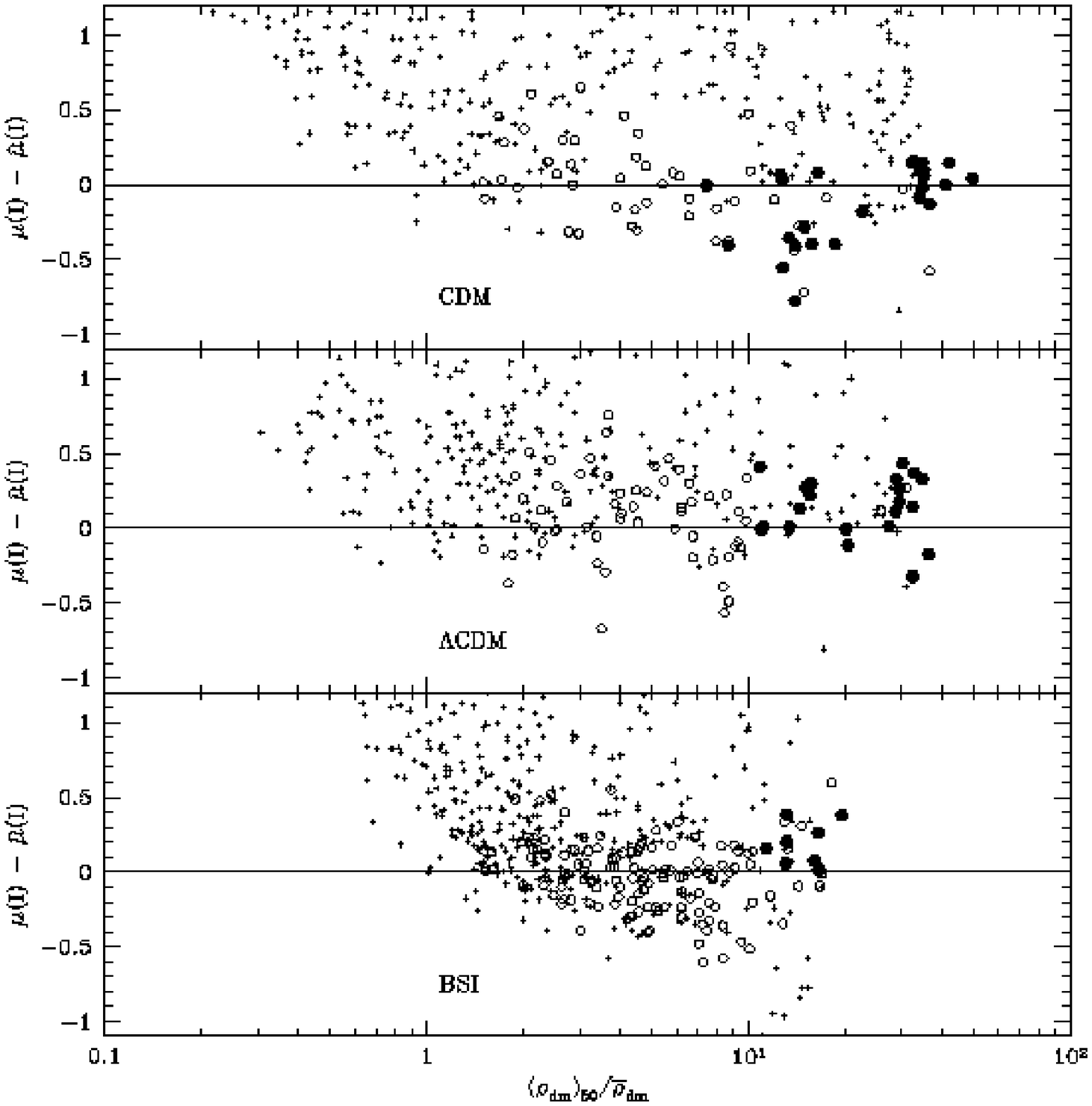,height=\ptamanio}}
\caption{Same as  in Fig. \protect\ref{fig:tfresb} but for the I band.}
    \label{fig:tfresi}
\end{figure}

Hence, the universality of the Tully-Fisher (TF) relation 
remains an open question.  If present, environmental bias in distance
determination could have important consequences for mapping the
large-scale velocity field.  In view of the importance of this relation
in cosmology and the 
difficulty of ascertaining the universality by empirical studies
(which must rely on adequate statistics), a theoretical prediction of 
invariance or expected bias would be quite useful.  

Using the numerical galaxies formed in our simulations, we have derived
 theoretical
relations analogous to the observed Tully-Fisher (TF) relations in
various photometric bands, for the three different cosmological models
We found that the observed slopes,
zero-points, and scatter of the TF relations are reproduced with
reasonable accuracy by models with a cosmological constant
 ($\Lambda $CDM) or models with broken scale invariance (BSI) , while
standard unbiased CDM leads to different slopes and/or zero points of
the relation \cite{yo,nosotros}.

In order to derive a theoretical TF relation, it is necessary
to model the circular velocity of our {\em numerical} galaxies. 
Unfortunately, the spatial resolution of 39 kpc does not permit an accurate
estimate of rotational velocities directly from the
simulation data.  Therefore, we devised  an operational
procedure for assigning ``rotational velocities",
defined by the depth of the gravitational potential: $v_{\rm
grav}=\sqrt{GM/r}$, where $M$ is the total mass of a galaxy within its
assigned radius $r$ (1 or 2 cells, i.e. 39 or 78 kpc). 
We have tested this procedure for  internal consistency with other indicators 
as reported in \cite{nosotros}.  In addition, we find 
that the simulated TF relation for our galaxies is stable with respect to  
variations in the details of the halo finding
algorithm used and  changes in the spatial resolution of the simulations.  The 
resolution check was carried out by comparing the $256^3$  and $128^3$
$\Lambda$CDM simulations reported in Section \ref{subsec:resolution}). Hence, 
we can be reasonably confident that our results primarily reflect the implications of
our model for galaxies rather than the details of the numerical procedures used here.
 
In this section, we continue our study of environmental effects
by searching for a significant dependence of the luminosity-line width relation on
environment  in a statistical sample of 
halos obtained in our simulations.  More precisely, we have performed a stepwise linear regression 
of luminosity on rotational velocity, color index, 
and environment (characterized as above by the ambient overdensity)

We first consider 
residuals to the best fits to the TF relation in the B and I bands
(i.e., $\hat{M}$): The residuals of the B-band and, to a lesser degree, the R-band
and I-band TF - relations were found to be correlated with color
index [$CI= (B-I)$ or $CI =(U-B)$] according to a relation of the
form $\mu(M) \equiv  M -\hat{M}=a CI + b$ , 
in agreement with  Giraud \cite{gir},
who   has found evidence for a color dependence of the
difference in distance moduli derived from the TF - relation
in two different bands.  The fit parameters $a$
and $b$ as well as the correlation coefficients $r$ are listed in
Table (\ref{table2}).  Note that only galaxies with $ B < -16$ are included
in the fit, as before.

A hypothesis test of significant regression was performed
by constructing the F--statistic for every fit.  The null hypothesis
$a=0$ was ruled out at the 99 \% confidence level for
all fits. As expected, the color dependence is least pronounced
(but still significant) in all scenarios in the I - band.
The slopes of the $\mu(M)$ vs. CI - relation are different for different
cosmological scenarios.  In all scenarios, the slope
of the relation is positive.  Thus, the trend is
for bluer galaxies to be brighter than predicted by the uncorrected TF - relation.

Having thus obtained a color-corrected TF-relation, we now compute the residuals with respect to the
corrected relation in these two bands and search for a relationship between these residuals and 
the ambient overdensity $<\rho_{dm}>_{0.5}/\bar{\rho}_{dm}$ as defined earlier. The regressions 
are performed separately for each scenario using all galaxies in the corresponding catalog.
After removing the color trend as above, regression on
the remaining residuals was thus performed according to the model
$\mu(M) -\hat{\mu}(M) =c \log (<\rho_{dm}>_{0.5}/\bar{\rho}_{dm} )
+ d $.  (Here  
$\hat{\mu}(M)$ is the residual predicted using
the color regression as above.)
 In Fig.
(\ref{fig:tfresb}) and (\ref{fig:tfresi}) the  B and I band residuals
$\mu(M) -\hat{\mu}(M)$  is plotted as a function of the 
ambient overdensity $<\rho_{dm}>_{0.5}/\bar{\rho}_{dm}$ of dark
matter.  We have also applied the hypothesis test of regression by
means of the F-statistic to  these plots.  Within the dynamical
range studied here, there was no  environmental  effect ($c=0$) in the TF
 at the 99 \% or even 95 \% confidence levels in BSI or $\Lambda$CDM.  
However, at the 99 \% confidence level there was an environmental
effect in the CDM sample: in the I-band, we find $c=-0.21$, $d=0.18$ and
in the B-band, $c=-0.36$, $d=0.33$.
Qualitatively similar results are obtained if $\mu(M)$
is modeled (residuals without color correction)
instead of $\mu(M) -\hat{\mu}(M) $.

Our results for the $\Lambda$CDM and BSI scenarios are consistent
with the usual view that the TF - relation is unbiased
with respect to environment.  However, despite the fact that we find 
a statistically significant environmental
dependence of the residuals on environment in our CDM sample, it would
not be correct to infer that the TF -relation for spirals
would have environmental bias in a CDM universe:  
the entire statistical effect could be due to inclusion of the  ``ellipticals'', classified
here on the basis of colors, as was discussed in Section
\ref{sec:mfenv}. Indeed, once the  ``elliptical'' galaxies are removed from the fit,  we find,
   at a   99\% confidence level, no  environmental effect on the TF
   relation for CDM either.

\section{Discussion and conclusions}
\label{sec:conclu}

We have presented  results of cosmological 
hydrodynamical simulations incorporating a multiphase model of the 
ISM, radiative cooling,  
 star formation, metal enrichment, and supernova feedback. 
The simulations have been performed for three cosmological 
scenarios: CDM, $\Lambda$CDM and BSI.

 The depth of modeling of the physical processes in the baryonic
 component  has permitted 
 us  to study the motion and evolution of gas inside and around dark
 matter halos. As described in {\S} \ref{sec:baryoncomp},
the gas flow is extremely complicated:  While we clearly see accretion
shocks at large distances from halos (up to 1 Mpc), the flow inside the
shock cannot be treated as spherical. A fraction of the gas flows along
filaments. It penetrates very deeply inside the ``caverns'' which are
produced by accretion shocks. We frequently found outflows of gas
inside caverns.  These outflows (``chimneys'') are not directly 
related with the supernova feedback, but in some cases are enhanced by it.
The distribution of the gas is very much
affected by the short-scale processes related with the star-formation
and its back-reactions to the surrounding gas. In particular,
metal enrichment  change the position of the accretion shock. 
 The higher  cooling rate  for a metal  enriched gas  causes it to 
 produce more cold clouds and weaker pressure gradients.
 Therefore, star formation   is more efficient,
 and galaxies become brighter and bluer. 

We have also  analyzed  the effects of environment on 
 the observational properties of our
 numerical galaxies, such as morphologies (as characterized by colors), 
 SFR and  Tully-Fisher relation.  
We can summarize the main results as follows:
\begin{itemize}
\item  The
percentages of galaxies  with different  morphologies 
differ markedly from one scenario
to another.  Concretely, the CDM simulations  produce a considerable population
of elliptical galaxies ($\sim 13$\%) with red colors and a dearth of
irregulars ($\sim 3$ \%) at the range of densities (in a 0.5 Mpc scale)
 studied here,
corresponding to the typical environments of loose groups and field
galaxies.  $\Lambda$CDM simulations produce fewer red ellipticals, ($\sim 2$ \%)
while BSI produces very blue galaxies dominated by spiral ($\sim 80$\%)
and irregular ($\sim 20$\%) types. Differences between the morphology
distributions produced in different scenarios persist even if the
ambient density is held constant.
\item The SFR history for massive halos ($M>3\times 10^{11} M_\odot$)
  exhibit a drop from $z\simeq1$ to $z=0$. This drop in the SFR
  is more pronounced  in halos located in higher density
  environments. But, for the same range of environment densities, the
  drop is larger for halos found in CDM simulations than  in
  $\Lambda$CDM. On the contrary, the SFR history of less massive halos 
   ($M<3\times 10^{11} M_\odot$)
does not  exhibit a clearly  defined correlation with the ambient density  at present epoch.

\item The Tully-Fisher relation for  the galaxies in $\Lambda$CDM  and BSI
  simulations  does not depend on environment, with a 99\% 
  confidence level. For CDM, we find a statistically significant
  environmental dependence of the TF-relation, but this effect is due
  to the inclusion of ``ellipticals'', that are 
 produced in this model. When 
 the ``ellipticals'' are removed from the  sample, we find 
no environmental effect, at a 99\%  confidence level as in the other
models.
\end{itemize}

We conclude that the ambient density is {\bf NOT} the
{\em only} factor defining  observable galaxy properties.  The
merging history, tidal encounters, and interactions with the
surrounding gaseous medium could  be important in determining the properties
of galaxies \cite{balland} located in regions of similar ambient
density.  Our simulation technique provides a point of departure for
such a study, since a representative sample of possible initial
conditions can be generated by different realizations.

We have also shown that a proper description of 
physical processes in the baryonic component is important in 
modeling the
formation and evolution of galaxies and in predicting their observational
properties.  We are still far from a complete understanding, let alone
modeling, of the physical processes that govern the formation and
evolution of galaxies.  Nonetheless, the YK$^3$ code takes into account
mechanisms corresponding to physical processes which according to
numerous lines of evidence are most important in the dynamics of
galaxies.  The modeling of these processes incorporates the available
physical and astrophysical knowledge to the depth possible consistent
with computational limitations and the need to minimize the number of
degrees of freedom in the parametrization.  As a general principle,
predictive power is strongly improved by parsimonious use of fitting
parameters, and indeed the ability to make predictions is the
difference between modeling and curve fitting.

Improvements in numerical resolution would certainly be useful in order
to extend our results of the effects of environment to larger-scale
structures.  Clearly, on the basis of the present results one cannot
exclude additional environmental effects that could result from the
existence of structures on scales beyond our current box size (5 Mpc).
Nevertheless, we can conclude that our numerical model for galaxy
formation, which includes the most relevant physical processes for the
baryonic component, is capable of reproducing most of the observational
trends of real galaxies. We have seen that for a reasonable expenditure
of computational resources, it is possible to study problems such as
environmental effects while including hydrodynamics. Hence, the
combination of hydrodynamical simulations with modeling of the baryonic
component constitutes a very useful and powerful tool for investigating
the complex phenomena of galaxy formation.


\newcommand{\AJ}[3] {~ #1 ~ApJ~  { #2}~ #3}
\newcommand{\AJL}[3]{~ #1 ~ApJ~  { #2}~ #3}
\newcommand{\AJS}[3] {~ #1 ~ApJS~  { #2}~ #3}
\newcommand{\AsA}[3] {~ #1  ~A\&A~  { #2}~ #3}
\newcommand{\MN}[3] {~ #1  ~MNRAS~  { #2}~ #3}

\end{document}